\documentstyle[emulateapj,apjfonts]{article}

\def\ROSAT{ROSAT}

\def\EXOSAT{EXOSAT}

\def\arcsec{\ifmmode '' \else $''$\fi}
\def\arcmin{\ifmmode ' \else $'$\fi}
\def\arcsecpoint{\ifmmode ''\!. \else $''\!.$\fi}
\def\arcminpoint{\ifmmode '\!. \else $'\!.$\fi}
\def\cc{\ifmmode {\rm cm}^{-3} \else cm$^{-3}$\fi}
\def\cl{\ifmmode {\rm cm}^{-2} \else cm$^{-2}$\fi}
\def\micron{\ifmmode \mu{\rm m} \else $\mu$m\fi}
\def\kms{\ifmmode {\rm km\,s}^{-1} \else km\,s$^{-1}$\fi}
\def\Hubble{\ifmmode {\rm km\,s}^{-1}\,{\rm Mpc}^{-1} 
	\else km\,s$^{-1}$\,Mpc$^{-1}$\fi}
\def\ergsec{\ifmmode {\rm ergs\;s}^{-1} \else ergs s$^{-1}$\fi}
\def\ergscm{\ifmmode {\rm ergs\,s}^{-1}\,{\rm cm}^{-2}
	  \else ergs\,s$^{-1}$\,cm$^{-2}$\fi}
\def\ergscmA{\ifmmode {\rm ergs\,s}^{-1}\,{\rm cm}^{-2}\,{\rm \AA}^{-1}
	  \else ergs\,s$^{-1}$\,cm$^{-2}$\,\AA$^{-1}$\fi}
\def\ergscmHz{\ifmmode {\rm ergs\,s}^{-1}\,{\rm cm}^{-2}\,{\rm Hz}^{-1}
	  \else ergs\,s$^{-1}$\,cm$^{-2}$\,Hz$^{-1}$\fi}
\def\Msun{\ifmmode M_{\odot} \else $M_{\odot}$\fi}
\def\Lsun{\ifmmode L_{\odot} \else $L_{\odot}$\fi}
\def\qo{\ifmmode q_{0} \else $q_{0}$\fi}
\def\Ho{\ifmmode H_{0} \else $H_{0}$\fi}
\def\lya{Ly$\alpha$}

\def\civ{C\,{\sc iv}}

\newcommand {\etal}{et al.}

\newcommand{\ovi}{O~{\sc vi}}

\lefthead{KRISS ET AL.}
\righthead{FAR-UV PEAK OF 3C~273}

\begin{document}
\submitted{To appear in the 1999 December 20 issue of the Astrophysical Journal}

\title{The Ultraviolet Peak of the Energy Distribution in 3C~273:\\
Evidence for an Accretion Disk and Hot Corona\\Around a Massive Black Hole}

\author{
Gerard A. Kriss\altaffilmark{1,2}, Arthur F. Davidsen\altaffilmark{2}, Wei Zheng
\altaffilmark{2}, and Geunho Lee\altaffilmark{2,3}
}
\altaffiltext{1}{Space Telescope Science Institute,
3700 San Martin Drive, Baltimore, MD 21218; gak@stsci.edu}
\altaffiltext{2}{Center for Astrophysical Sciences, Department of Physics and
Astronomy, The Johns Hopkins University, Baltimore, MD 21218--2686; afd@pha.jhu.
edu, zheng@pha.jhu.edu}
\altaffiltext{3}{Radio Research Laboratory, 370-9, Sinpilli, Seolseong-Myun,
Ichon, Kyoungki-do, 467-880 Korea; lgh@solaradio.rrl.go.kr}

\begin{abstract}

We present absolutely calibrated far-ultraviolet spectrophotometry
of the quasar 3C~273 covering the 900--1800 \AA\ range.
Our $\sim$ 3 \AA\ resolution spectra were obtained with the
Hopkins Ultraviolet Telescope during the Astro-1 mission in
December 1990 and during the Astro-2 mission in March 1995.
Both spectra exhibit a change in slope near the Lyman
limit in the quasar rest frame.
At longer UV wavelengths, the continuum has a power-law index of
0.5--0.7, while shortward of the Lyman limit it is 1.2--1.7.
The energy distribution in $\nu f_{\nu}$ therefore peaks close to
the quasar Lyman limit.
The short wavelength UV power law extrapolates well to match the
soft X-ray excess seen in simultaneous observations with the
Broad-Band X-ray Telescope and nearly simultaneous ROSAT observations.

The general shape of the broad-band spectrum of 3C~273 is consistent with
that of an optically thick accretion disk whose emergent spectrum has been
Comptonized by a hot medium.  Our UV spectrum is well described by a
Schwarzschild black hole of $7 \times 10^8~\rm \Msun$ accreting matter
at a rate of $13~\rm\Msun~yr^{-1}$ through a disk inclined at 60 degrees.
Superposed on the intrinsic disk spectrum is an empirically determined
Lyman edge of optical depth 0.5.  The Comptonizing medium has a Compton
parameter $y \approx 1$, obtained with an optical depth to electron scattering
of unity and a temperature of $4 \times 10^8~\rm K$.

This overall shape is the same as that found by Zheng et al. and Laor et al.
in their UV and X-ray composite spectra for quasars, giving physical validity
to the composite spectrum approach.
When combined with those results, we find that the generic ionizing
continuum shape for quasars is a power law of energy index 1.7--2.2,
extending from the Lyman limit to $\sim$1 keV.
The observational gap in the extreme ultraviolet for these combined data
describing the quasar continuum shape is now only half a decade in frequency.

\end{abstract}


\section{Introduction}

It has long been known that the energy distribution of quasars, expressed 
as $\nu L_{\nu}$, must have a peak somewhere in the far-ultraviolet to
soft X-ray region of the spectrum. Shields (1978)\markcite{Shields78}
first suggested that the optical and near UV flux of quasars might be due
to the Rayleigh-Jeans portion of a black body spectrum whose peak lies in
the unobserved extreme UV region, arising from an optically thick accretion
disk surrounding a massive black hole (\cite{LB69}),
and dubbed the ``big blue bump.''
This idea was then further developed and applied to UV observations of quasars
obtained with IUE (\cite{MS82}; \cite{Malkan83}). It was
subsequently found that the X-ray spectra of quasars also showed a ``soft
excess'' of flux, above the extrapolation of the relatively flat power-law
that fit the spectrum at energies above 2 keV
(\cite{TP88}; \cite{Masnou92}), and 
which might arise from the Wien portion of the same thermal
spectrum. In addition to whatever light this spectral region might shed on
the black-hole accretion-disk model for quasars, it is also a fundamental
input to studies of the photoionization of the broad emission line clouds
in quasars (\cite{KK88}) as well as the photoionization of the
intergalactic medium (\cite{HM96}).
Consequently there has been widespread interest in
determining the detailed shape of the spectrum of quasars between the
optical ($10^{15}$ Hz) and the soft X-ray ($10^{17}$ Hz), and in particular,
establishing more precisely where in the extreme UV the peak of the big
blue bump is located (\cite{Arnaud85}; \cite{Bechtold87}; \cite{OGW88}).

\begin{deluxetable}{ccccc}
\tablewidth{4.3in}
\footnotesize
\tablecaption{Observation Log\label{tbl-1}}
\tablehead{
&\colhead{Instrument}
& \colhead{Date} & \colhead{Exposure} &
\colhead{Comment}\nl
&& \colhead{Start Time (GMT)} & (sec) &
}
\startdata
Astro-1&HUT & 1990 December 05  & 1251 & \nl
& & 23:28:24 & &  \nl
   &         & 1990 December 09 & 2138 & \nl
& & 17:23:54 & &  \nl
& BBXRT & 1990 December 09   & 1598   &   \nl 
  &    & 17:37:01 & & \nl
&IUE SWP40331 & 1990 December 14  &  1800 & \nl
&~~~ SWP40333 &  & 1500 & \nl
&~~~ LWP19411 &  & 1800 & \nl
&~~~ LWP19412 &  & 1800 & \nl
&~~~ SWP40391 & 1990 December 19  & 1800 & \nl
&~~~ SWP40392 &  &  1800 & \nl
&~~~ LWP19447 & & 1800 & \nl
&~~~ LWP19448 & & 1620 & \nl
&\ROSAT\ & 1990 December 18--21 & 497 & \nl
&KPNO 2.1m & 1990 December 05 & 60 & \nl 
AA\nl
\tableline
Astro-2&HUT& 1995 March 12 & 1180 & Flux correction 1.02 \nl
           & & 06:05:03  & & \nl
           & & 1995 March 15 & 1304  & Flux correction 1.05\nl
           & & 07:11:43 & & \nl
\enddata
\end{deluxetable}

The quasar 3C~273 provides an excellent candidate for addressing
this problem because of its low redshift (z=0.158) and its high flux at
both UV and X-ray wavelengths. It was the first quasar detected in X-rays
(\cite{Bowyer70}) and also the first quasar observed in the
far-ultraviolet (\cite{DHF77}), both detections having
been achieved with sounding rocket experiments before the advent of
effective satellite observatories.

At X-ray energies the spectrum of 3C~273 in the 2--10 keV range is
well fit by a power-law, $F_{\nu} \propto \nu^{-\alpha}$, with energy index
$\alpha \approx 0.5$ (\cite{Worrall79}), but observations at lower energies
with \EXOSAT\ and {\it Ginga} (\cite{Turner85}, \cite{Courvoisier87},
\cite{Turner90}), {\it Einstein} (\cite{WE87},
\cite{Turner91}, \cite{Masnou92}),
and \ROSAT\ (\cite{Staubert92}; \cite{LMP95},
hereafter LMP95) have established the existence of a soft excess.
Reporting on simultaneous observations of 3C~273 in
1990 December with \ROSAT\ (0.1 - 2 keV) and {\it Ginga} (2--10 keV), Staubert
et al. (1992)\markcite{Staubert92} obtained a good fit with the sum of two
power laws, with $\alpha = 0.56$ dominating at high energies and
$\alpha = 2.5$ at low energies.
However, the soft component could also be modeled with a black body spectrum or
with thermal bremsstrahlung (\cite{Staubert92}).
More extensive observations of
3C~273 with \ROSAT\ have been reported by LMP95\markcite{LMP95}.
They established that ``two
physically distinct emission components are present'', with the spectrum
best modeled by a combination of two power-laws with absorption by the
Galactic column of gas, $N_H = 1.84 \times 10^{20}$ \cl.
Constraining the harder component to have $\alpha_h = 0.5$ as observed
by \EXOSAT\ and {\it Ginga}, the soft
component was found to be significantly steeper, with $\alpha_s = 1.7$ for
most of the observations. Attempts to fit the soft component with various
other models, including blackbody and bremsstrahlung radiation were found
unacceptable (LMP95\markcite{LMP95}).

	At ultraviolet wavelengths the first determination of the spectral
index of 3C~273 (\cite{DHF77}) gave $\alpha = 0.6$ over the
range from the optical to Lyman $\alpha$, with a broad bump around 3500 \AA,
perhaps associated with Balmer continuum and Fe {\sc ii} line emission 
(\cite{Baldwin75}), that has
sometimes been called the ``small blue bump''. It was also pointed out by
Davidsen et al. that the spectral index must increase to $\alpha  > 1.2$
somewhere near or above the Lyman limit in order to agree with the X-ray
observations. Subsequent extensive observations with IUE give
$\alpha = 0.66$ between 3100 and 1250 \AA\ (\cite{ZM93}). 
A very low resolution observation with the Voyager UV
Spectrometer in the 900--1200 \AA\ band suggested the possibility of a spectral
break at about 1000 \AA\ (\cite{Reichert88}), perhaps associated with the
Lyman edge.  Recent observations in the 900--1200 \AA\ region with ORFEUS
imply a turnover in the spectrum of 3C~273 near the IUE/ORFEUS transition
wavelength at about 1200 \AA\ (\cite{Appenzeller98}).

	Here we report the results of observations of 3C~273 in the 900--1800
\AA\ region made with the Hopkins Ultraviolet Telescope (HUT) on both the
Astro-1 and Astro-2 space shuttle missions. The Astro-1 far-ultraviolet
data are accompanied by contemporaneous observations at optical,
ultraviolet, and soft and hard X-rays, yielding a quasi-simultaneous
broad-band spectrum for this quasar extending over more than 3 decades of
frequency. The HUT data alone show a definite spectral break, with a peak
in the energy spectrum $\nu L_{\nu}$ at about 920 \AA\ in the rest frame.  The
best-fit power-law at shorter wavelengths from 920 \AA\ to 787 \AA, when
extrapolated, nearly matches the simultaneous soft X-ray spectrum. The
Astro-2 HUT data are of even higher quality than those obtained on Astro-1
and yield similar results, confirming that the peak of the ``big blue
bump'' has now definitely been seen in 3C~273. A comparison of these data
with a composite quasar spectrum in the ultraviolet (\cite{Zheng97}) and
the X-ray bands (\cite{Laor97}), suggests strongly that a peak in the
energy spectrum of all quasars occurs near the Lyman limit, and that the
ionizing continuum from the Lyman limit to 1 keV is a power-law 
of $\alpha = 1.7 - 2.2$.

\begin{figure*}[t]
\plotfiddle{"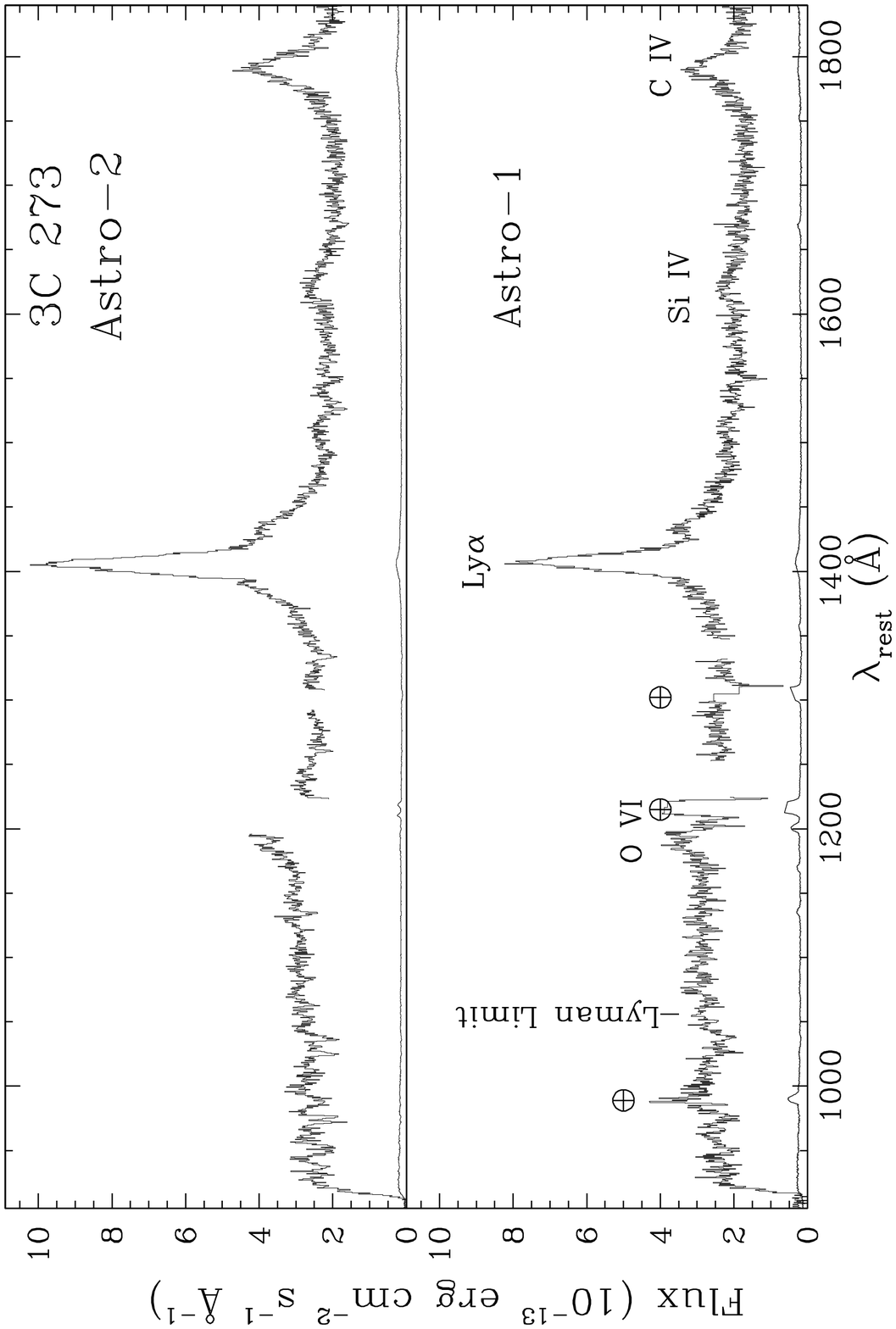"}{4.65 in} {-90}{70}{70}{-275}{390}
\caption{
3C~273 spectrum (histogram) from 900 to 1820 \AA\ as observed with HUT
on Astro-1 (lower panel) and Astro-2 (upper panel). Pixels are about 0.51 \AA\
and the resolution is about 3 \AA. Curves at the bottom of each panel
give the values of the error array produced by the standard HUT reduction
process. Strong peaks in the $1 \sigma$ error array for Astro-1 correspond to
strong geocoronal emission lines (airglow) for this observation, obtained
mostly in daylight. Airglow lines have been subtracted using templates
developed from blank sky observations, and their positions are marked with
\earth. The Astro-2 spectrum was obtained at night, resulting in much weaker
airglow. Emission and absorption lines are marked.
\label{fig1.ps}}
\end{figure*}

\section{Observations}

The Hopkins Ultraviolet Telescope (HUT) 
incorporates a 0.9-m primary mirror and a prime-focus Rowland
circle spectrograph as described by Davidsen et al. (1992)\markcite{Davidsen92}
and Kruk et al. (1995)\markcite{Kruk95}.
First-order spectra recorded by the photon counting detector cover
the 820--1840 \AA\ spectral range with a sampling of about 0.52 \AA\ per pixel 
and
a point source resolution of about 3 \AA. Absorption by interstellar hydrogen
limits the observed spectra to wavelengths longer than the Lyman limit at
912 \AA\ (787 \AA\ in the quasar rest frame).

HUT was used to observe 3C~273
on the Astro-1 space shuttle mission in 1990 December and again on the
Astro-2 mission in 1995 March. The performance and calibration of the
instrument are described for Astro-1 by Davidsen et al.
(1992)\markcite{Davidsen92} and Kruk et al. (1997)\markcite{Kruk97} and for
Astro-2 by Kruk et al. (1995\markcite{Kruk95}, 1999\markcite{Kruk99}).
Because substantial changes to HUT's
performance were made between the two flights, the results of the
spectrophotometric observations may be regarded as independent
measurements, effectively produced by two different instruments. The
Astro-1 observations were (necessarily) conducted primarily during daylight
portions of the shuttle orbit, where there is substantial contamination by
geocoronal emission lines that must be subtracted in the data reduction
process.
The Astro-2 observations, however, were obtained mostly during
night-time portions of the orbit, where the geocoronal contamination is
greatly reduced. Only the night portion of the Astro-2 data is used. 
Details of the observations are given as an observation log in
Table~1.

For Astro-1 there are several nearly simultaneous observations that
can be used to determine the broad-band spectrum of 3C~273. Also mounted on
the shuttle with HUT was the Broad Band X-Ray Telescope (BBXRT)
(\cite{Serlemitsos92}), which observed 3C~273 in the 0.3--12 keV band
with a resolution of about 100 eV. \ROSAT\ observations in the 0.1--2.4 keV
band were obtained less than two weeks after the Astro-1 observations
(\cite{Staubert92}), and {\it Ginga} observations in the 2--20 keV band
were also obtained during 1990 December (\cite{Staubert92}). 
In the ultraviolet, IUE spectra in both
the short and long wavelength cameras were obtained between one and two
weeks after the HUT observations, and these have been extracted from the
IUE archive. Finally, an optical spectrum of 3C~273 was kindly obtained for
us by R. Green on the KPNO 2.1 m telescope during the Astro-1 mission.  All
these contemporaneous measurements are listed in the observation log in
Table~1.

The HUT observations of 3C~273 from Astro-2 in 1995 March are also
listed in Table~1.
Unfortunately, we have been unable to find any nearly
simultaneous observations at other wavelengths for this epoch.

\section{Results from the HUT Observations}

\subsection{Observed Spectra}

The two HUT observations of 3C~273 from Astro-1 were summed and then
processed and reduced to an absolute flux scale following our standard
procedures as described by Kruk et al. (1997)\markcite{Kruk97}.
For Astro-2 the two
observations were reduced separately because of the variation of the
instrument sensitivity during the mission (\cite{Kruk95}), and the
resulting spectra were then averaged, weighted by their exposure time,
to yield the final spectrum presented here.
As noted in Table~1,
the Astro-2 data had slight flux corrections of 2--5\%
applied due to light loss at the slit induced by pointing errors.
The results for both missions are displayed in Fig.~1.

It is worth emphasizing that the Astro-1 and Astro-2 spectra were
obtained, reduced, and calibrated under markedly different circumstances.
The Astro-1 data were obtained with an instrument of relatively modest
sensitivity, almost entirely during orbital day when there is substantial
contamination by geocoronal emission lines that must be subtracted, and
were absolutely calibrated by reference to an observation of the DA white
dwarf G191-B2B using a model atmosphere computed for this star by P. Bergeron
(\cite{Davidsen92}). The Astro-2 data, on the other hand, were obtained
with an instrument of much higher sensitivity, entirely during orbital
night with minimal geocoronal contamination, and were absolutely calibrated
by reference primarily to observations of the DA white dwarf HZ43 using a
model atmosphere computed by D. Finley using D. Koester's model atmosphere
codes (\cite{Kruk95}).
In spite of these major differences, the resulting
spectra in Fig.~1 are remarkably similar.
Of course, 3C~273 is known to vary, so the close agreement between
observations made more than four years apart is fortuitous.
Indeed there are clear changes observed both in the continuum and
in the emission line strengths (discussed below).
However, both spectra show a change of continuum slope near the
Lyman limit in the rest frame of the quasar,
which is further discussed in the next section.

\subsection{Empirical Models for the Spectra}

We first remove the airglow lines from the spectra. We use a geocoronal \lya\
template that is derived from blank fields and scaled by the exposure time, 
and subtract it from the HUT spectrum of 3C~273. Other airglow lines are
fitted and removed, but significant residuals still exist in the regions around
the strongest airglow lines at 1216, 1302 and 989 \AA. 

We use the IRAF task {\it specfit} (\cite{Kriss94}) to fit both the Astro-1 and
Astro-2 spectra with various components for the continuum, emission lines
and absorption features.
We use the same overall model for both spectra, and identical fitting
windows spanning the wavelength intervals
912--1194, 1238--1287, and 1319--1820 \AA. 
For the Astro-1 data, we also model the remaining airglow residuals
near 989 \AA.

The intrinsic emission lines of
\lya, \civ\ and \ovi\ are modeled with dual Gaussians, and other emission  
components are modeled with single Gaussians.
We constrain many components of the fit by tying related parameters together.
The full-widths at half maximum (FWHM) of the weaker broad emission lines
({\sc S~vi}, {\sc C~iii}, {\sc N~iii}, {\sc O~i}, {\sc C~ii})
are all linked, and their wavelengths are linked to that of narrow Ly$\alpha$
by the ratio of their laboratory wavelengths.
The FWHM of broad {\sc O~vi} $\lambda$1034 is linked to broad {\sc C~iv}
$\lambda$1549.
The wavelengths and FWHM of the {\sc N~v} components are linked to the
corresponding {\sc C~iv} components.
We include one unidentified weak, broad emission feature at a rest
wavelength of 1074 \AA.  This is not He~{\sc ii} $\lambda$1085, and it is
approximately at the location of the unidentified feature also seen in
Faint Object Spectrograph (FOS) observations by Laor et al.
(1995)\markcite{Laor95}.

For the absorption lines, we start by including features identified in
previous, higher resolution observations (FOS: \cite{Bahcall91};
GHRS: \cite{Morris91}; ORFEUS: \cite{Hurwitz98}).
We then add additional features as required to obtain a good fit
to the HUT data.
As with the emission lines, we constrain many components of the fit
by tying the FWHM of the lines together in groups by wavelength regions
where the instrumental resolution is roughly the same.
All lines are assumed to be unresolved, except for the blend of
{\sc O~vi}, {\sc C~ii}, and {\sc O~i} around 1038 \AA,
the blended Si~{\sc ii} lines at 1192 \AA, and
the blended {\sc C~iv} doublet at 1549 \AA.

The wavelength region
below 950 \AA\ is heavily absorbed by the high-order lines of Galactic 
hydrogen. To model this, we use a set of absorption profiles, each including
50 Lyman-series lines, with a fixed column density of $1.8 \times 10^{20}$ \cl.
The absorption lines are modeled using Voigt profiles covering a range of
10--20 \kms\ in the Doppler parameter which are then convolved with the
instrument resolution, a Gaussian of 3 \AA\ FWHM.
For our extinction correction, we adopt $E(B - V) = 0.032$, as established
by Lockman \& Savage (1995)\markcite{LS95}, and use the extinction curve
of Cardelli, Clayton \& Mathis (1989)\markcite{CCM89}, with $R_V = 3.1$.
This value of the extinction is also consistent with the $E(B - V) = 0.03$
adopted by Lichti et al. (1995)\markcite{Lichti95} and
Appenzeller et al. (1998)\markcite{Appenzeller98}.

For both the Astro-1 and Astro-2 spectra, we obtain our best fit using a
broken power-law model for the continuum shape.
To ascertain the significance of the apparent spectral break near the Lyman
limit, we compare fits performed using a single power-law for the continuum
shape to a broken power-law.
The single power-law fits are summarized in Table~2,
and the broken power-law fits are summarized in Table~3.
For the Astro-1 data, the fits with a single power law do not produce
satisfactory results.
These fits are shown as smooth curves in Fig.~2
and Fig.~3.
Note that the single power-law fit
deviates significantly at wavelengths shortward of 1200 \AA.
Even allowing for reddening corrections higher than the
Galactic value, the fits are still poor, as shown in Table~2.
The spectral break in the sub-\lya\ region is too sharp to be accounted for by
reddening. 

\begin{center}
\small
{\sc TABLE 2\\
Single Power-Law Fits to 3C~273}
\vskip 4pt
\begin{tabular}{ccccc}
\hline
\hline
$E(B-V)$ & \multicolumn{2}{c}{Astro-1} & \multicolumn{2}{c}{Astro-2}\\
         & $\alpha$ & $\chi^2/dof$ & $\alpha$ & $\chi^2/dof$\\
\hline
0.01 & \phantom{$-$}0.94 & 1689/1544 & 1.16 & 1723/1501\\
0.02 & \phantom{$-$}0.81 & 1651/1544 & 1.00 & 1698/1501\\
0.032 & \phantom{$-$}0.62 & 1619/1544 & 0.79 & 1697/1501\\
0.04 & \phantom{$-$}0.49 & 1615/1544 & 0.67 & 1708/1501\\
0.05 & \phantom{$-$}0.34 & 1602/1544 & 0.50 & 1746/1501\\
0.06 & \phantom{$-$}0.18 & 1607/1544 & \nodata & \nodata\\
0.08 &           $-0.15$ & 1650/1544 & \nodata & \nodata\\
\hline
\end{tabular}
\end{center}

\begin{center}
\small
{\sc TABLE 3\\
Broken Power-Law Fits to 3C~273$\rm ^a$}
\vskip 4pt
\begin{tabular}{ccc}
\hline
\hline
Parameter & Astro-1 & Astro-2\\
\hline
$f_0 \rm ^b$ & $4.37 \pm 0.04$ & $4.41 \pm 0.02$ \\
$\lambda_0$ (\AA) & $1064 \pm 8$ & $ 1036 \pm 6$ \\
$\alpha_1$ & $1.69 \pm 0.16$&  $1.19 \pm 0.15$\\
$\alpha_2$& $0.46 \pm 0.03 $&  $0.74 \pm 0.02 $ \\
$\chi^2/dof$& 1584/1542 & 1692/1499\\
\hline
\end{tabular}
\vskip 2pt
\parbox{3.5in}{
\footnotesize
$\rm ^{a}$Assumes extinction fixed at $E(B-V)=0.032$.\\
$\rm ^{b}$Measured at the break wavelength $\lambda_0$, in units of
$10^{-13}$ \ergscm.
}
\end{center}

Although the Astro-2 spectrum is fairly similar to the Astro-1 data,
the continuum shape is noticeably less blue at wavelengths longward of \lya,
and the broad emission lines are significantly brighter.
One can see this most easily in the comparison of the fitted spectra
shown in Fig.~4.
Since the Astro-2 spectrum is not as blue,
the break at shorter wavelengths is not as dramatic, and a
single power law reddened at the nominal Galactic value fits only slightly
worse than the broken power law as one can see in the top panels
of Figures~2 and 3.
Thus, we conclude that a short wavelength spectral break is required by
our Astro-1 data, but not by the Astro-2 data.

\vbox to 7.65in {
\vbox to 14pt{\vfill}
\plotfiddle{"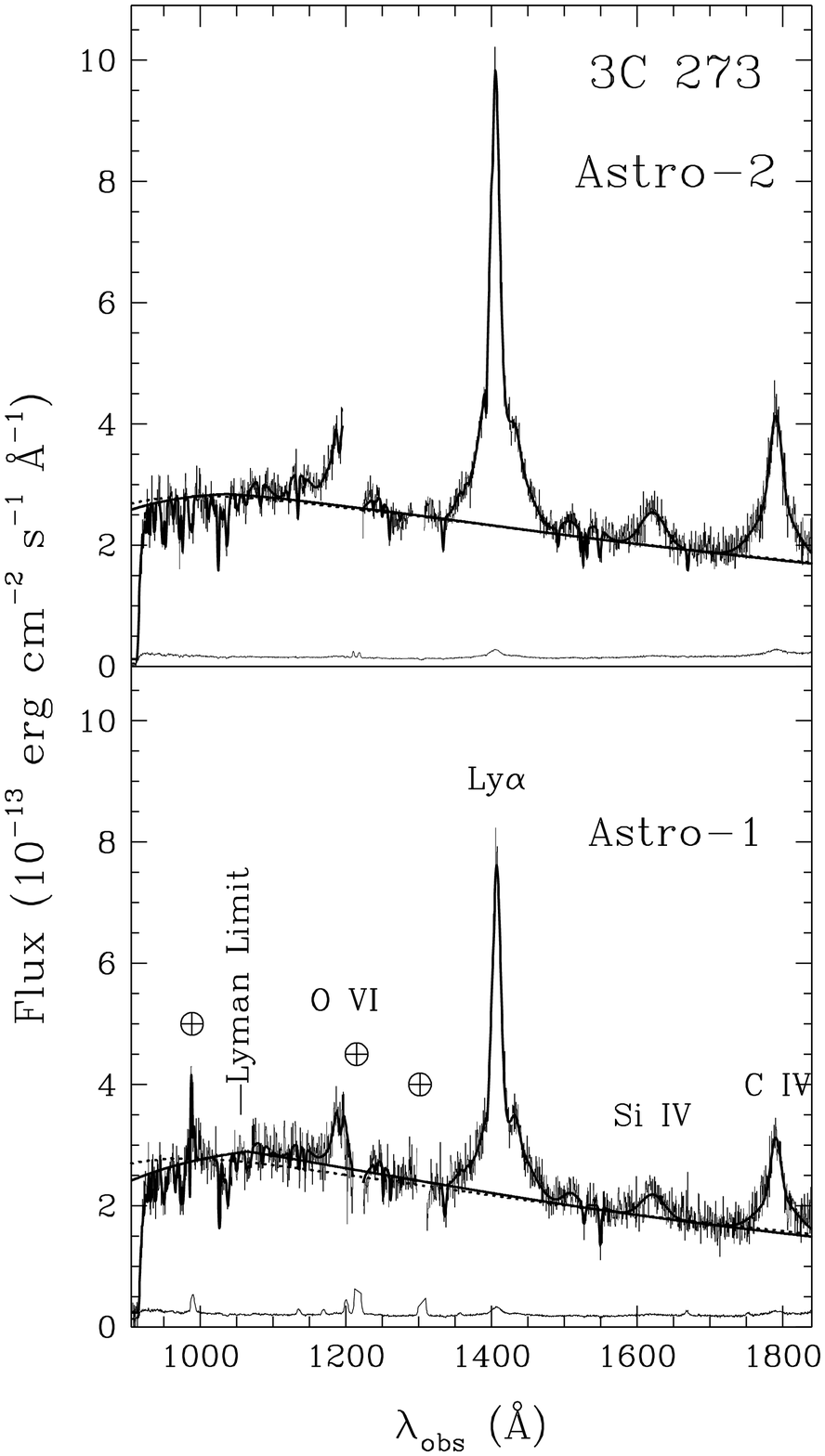"}{6.05 in} {0}{60}{60}{-182}{-10}
\parbox{3.5in}{
\small\baselineskip 9pt
\footnotesize
\indent
{\sc Fig.}~2.---
Observed spectra of 3C~273 with the best-fitting empirical models as
described in Section 3.2. The models include emission and absorption lines
and a continuum described either by a single power-law (dotted line) or by
a broken power-law (solid lines). Reddening with $E(B-V) = 0.032$ is applied
to the models. Broken power-laws provide the best fit for both spectra.
A single power-law gives an unacceptable fit for the Astro-1 spectrum;
for Astro-2 a single power-law is worse, but acceptable.
Parameters of the models are given in Tables~2--7.
\label{fig2.ps}}
\vbox to 14pt{\vfill}
}

As the broken power law continuum models give our best fits for both data sets,
we quote the parameters for the emission and absorption lines using these
continuum models.  Tables~4 and 5 give the observed
wavelengths, the observed
fluxes, and full-widths at half-maximum (FWHM) for the Astro-1 and Astro-2
observations of the emission lines, respectively.
Tables~6 and 7 list the observed wavelengths,
equivalent widths ($W_\lambda$), and FWHM for the absorption lines.
None of the line widths are corrected for the instrumental resolution.
For all absorption lines, the best-fit widths are consistent with the
instrumental resolution and should therefore be considered unresolved.
The error bars quoted on all parameters are determined from the error matrix
of the fit, and they represent the 1$\sigma$ confidence interval for a single
interesting parameter (\cite{Kriss94}).

\vbox to 7.1in {
\vbox to 14pt{\vfill}
\plotfiddle{"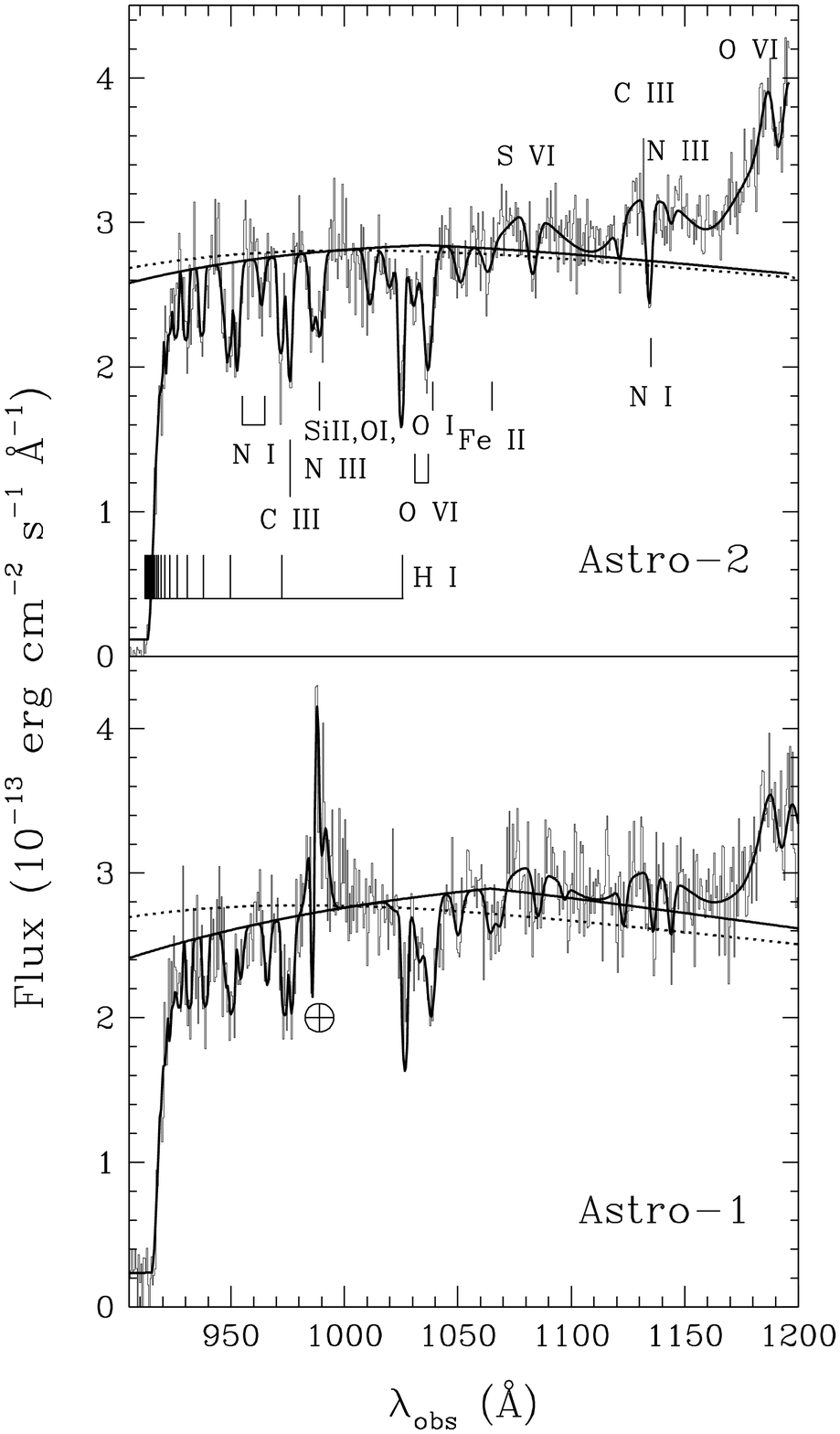"}{5.95 in} {0}{57}{57}{-177}{-10}
\parbox{3.5in}{
\small\baselineskip 9pt
\footnotesize
\indent
{\sc Fig.}~3.---
Same as Fig.~2 but showing the crucial 900--1200 \AA\ region at an
enlarged scale. Positions of the Lyman series of interstellar hydrogen
absorption lines are indicated. Series lines up to L7 are clearly seen in
the data. The continuum in this region is well-fit by the broken power-law
model (solid curves), while the best single power-law model (dotted) lies
above the data at the shortest wavelengths.
\label{fig3}}
\vbox to 14pt{\vfill}
}

\section{The Broad-band Spectrum of 3C~273}

The BBXRT data were obtained 
simultaneously with the Astro-1 HUT data, and we have retrieved them from the
National Space Science Data Center (NSSDC) archive. To obtain the unfolded 
X-ray spectrum, we use the data reduction package
{\it xspec} (\cite{Arnaud96}). The data points in channels 1--20 and 450--512
are excluded because of high uncertainties. We bin the data so that each bin
has at least 20 counts. Our preliminary fits with single power laws or
broken power laws yielded poor fits,
with a significant deviation 
around 0.6 keV. Done (1993)\markcite{Done93} suggests that this is a
blueshifted O {\sc viii} resonance absorption feature. 
Empirically modeling this feature as an absorption edge,
we use a dual power-law with fixed Galactic absorption to
model the X-ray spectrum:

$$ dN / dE = (f_1 E^{-\alpha_1} + f_2 E^{-\alpha_2})
{\rm exp}(-\tau(E/E_{th})^{-3}) {\rm exp}(-\tau(N_H,E)).$$

This model gives a good fit ($\chi^2 = 52.5$ for 52 binned data points
and 6 free parameters). 
The parameters
for our best fit are: the photon power-law index $\alpha_1 = 4.3 \pm 0.9$,
$\alpha_2 = 1.64 \pm 0.09 $; flux at 1 keV: $f_1 = 0.0081 \pm 0.0030$ and  
$f_2 = 0.017 \pm 0.0025$ $\rm photon\ s^{-1}\ cm^{-2}\ keV^{-1}$; and the 
absorption edge at $E_{th} = 0.51 \pm 0.03$ keV with an optical depth
$\tau = 1.17 \pm 0.31$.
This fit and the residuals are shown in Fig.~5.

\vbox to 4.8in {
\plotfiddle{"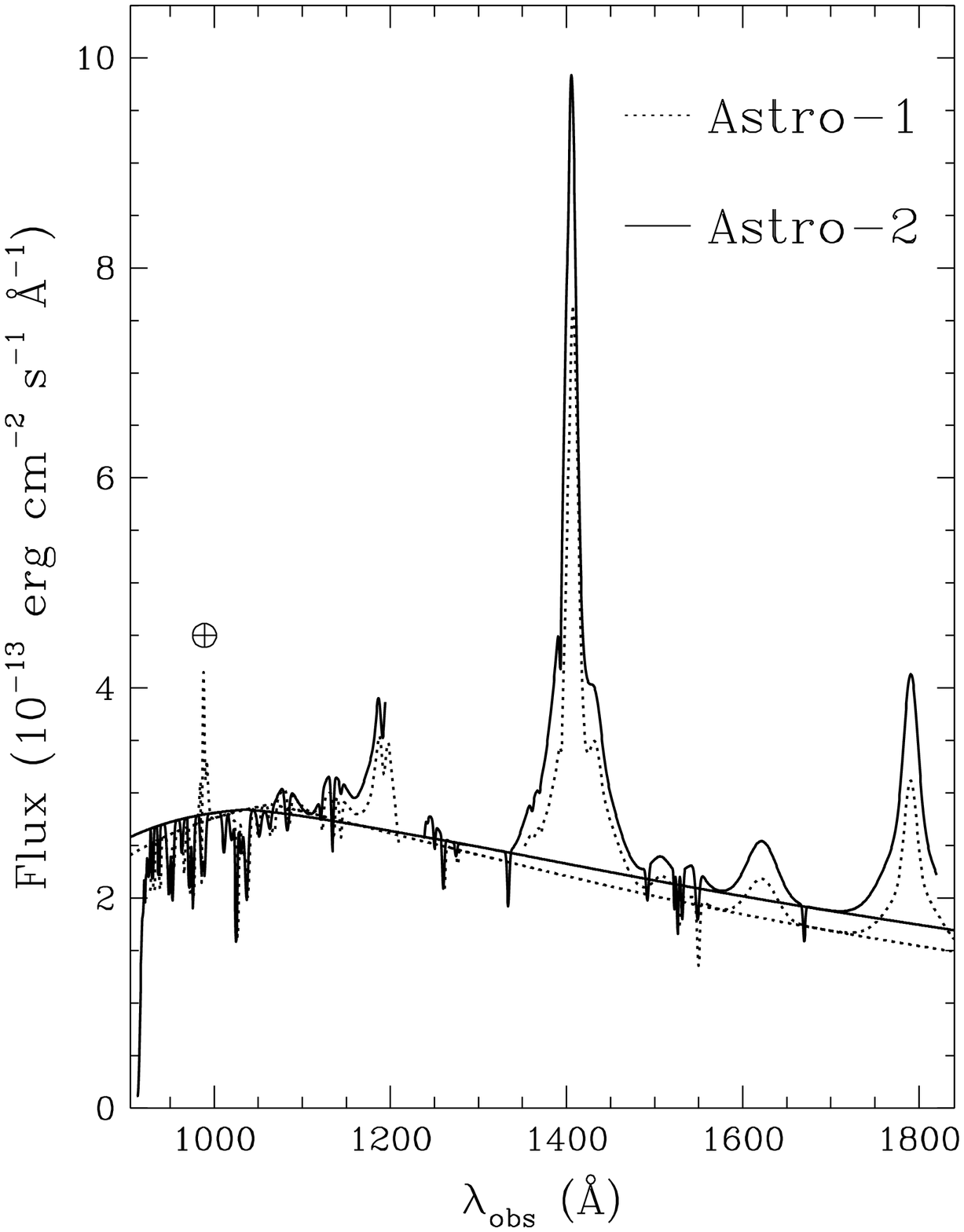"}{4.25 in} {0}{50}{50}{-162}{-60}
\parbox{3.5in}{
\small\baselineskip 9pt
\footnotesize
\indent
{\sc Fig.}~4.---
Direct comparison of the fits to the HUT Astro-1 (dotted lines)
and Astro-2 (solid lines) spectra.  Although the continuum fluxes are roughly
the same, there are clear differences in the line fluxes and continuum shapes.
\label{fig4.ps}}
\vbox to 14pt{\vfill}
}

The nearly simultaneous optical, ultraviolet, and X-ray data
collected in 1990 December during or shortly after the Astro-1 mission were
listed in Table~1 and are shown in Fig.~6.
The BBXRT and \ROSAT\ data shown in Fig.~6 are corrected for
Galactic absorption, using a Morrison \& McCammon (1983)\markcite{MM83} model
with $N_H = 1.8 \times 10^{20}$ \cl.
(Note that we show {\it unfolded} X-ray data in Fig.~6,
and that the plotted fluxes are model dependent.)
Also shown is the power-law that
best fits the short wavelength portion of the HUT spectrum as part of the
broken power-law fit described in Section 3.2.
This component has $\alpha = 1.7$.
Two additional power-laws are also shown as dashed lines with indices
of $1.7 \pm 0.36$, representing the
1-$\sigma$ errors from our fit added linearly to the uncertainty of
$\pm 0.2$ corresponding to a range in $E(B-V)$ of $\pm 0.01$,
which is our estimate of the uncertainty in the extinction for 3C~273.
It is clear from Fig.~6 and Fig.~2 that
the energy distribution of 3C~273 peaks within the HUT spectral range, a
little longward of the Lyman limit in the quasar rest frame, and that the
short wavelength power-law that fits the HUT data extrapolates well to
fit the soft X-ray data from \ROSAT\ and BBXRT.

There still remains a considerable gap of one decade in frequency
that cannot be filled for 3C~273, however, because the ultraviolet spectrum
cannot be extended past the Galactic Lyman limit at 787 \AA\ in the quasar 
rest frame, and the soft X-ray spectrum is severely attenuated by
interstellar absorption below 0.2 keV. Indeed this extreme ultraviolet gap
cannot readily be decreased in the spectrum of any single quasar, since
the higher redshifts that would allow the ultraviolet spectrum to be
extended will at the same time limit the observed X-ray spectrum to higher
rest-energies. We therefore turn to a composite quasar spectrum assembled
from many different objects to try to reduce the size of the extreme
ultraviolet gap.

In a previous paper (Zheng et al. 1997) we have presented the
results of an analysis of HST FOS spectra of 101 quasars, most with
redshifts from 0.33 to 1.5, with a few high redshift objects with $z$ up to
3.6. The composite quasar spectrum was also fit with a broken power-law,
the break occurring around 1050 \AA\ in the rest frame.
The short wavelength component was found
to have $\alpha = 2.0$ for the total sample (which is subject to unknown
selection effects), while the subset of 60 radio-loud quasars had $\alpha =
2.2$ and the subset of 41 radio-quiet quasars had $\alpha = 1.8$.

In the lower panel of Fig.~6 we plot the radio-loud composite
quasar spectrum from Zheng et al. (1997)\markcite{Zheng97} scaled to match the
flux observed for 3C~273 at 1000 \AA, near the break
point in both spectra. Also shown as a solid line (with a break at 2 keV)
is a composite quasar X-ray spectrum for radio-loud quasars from Laor et al.
(1997)\markcite{Laor97}, which has been scaled in flux to match our UV-optical
composite using the average $\alpha_{ox}$ of 1.445 from the sample of
Laor et al.

\vbox to 4.5in {
\vbox to 14pt{\vfill}
\plotfiddle{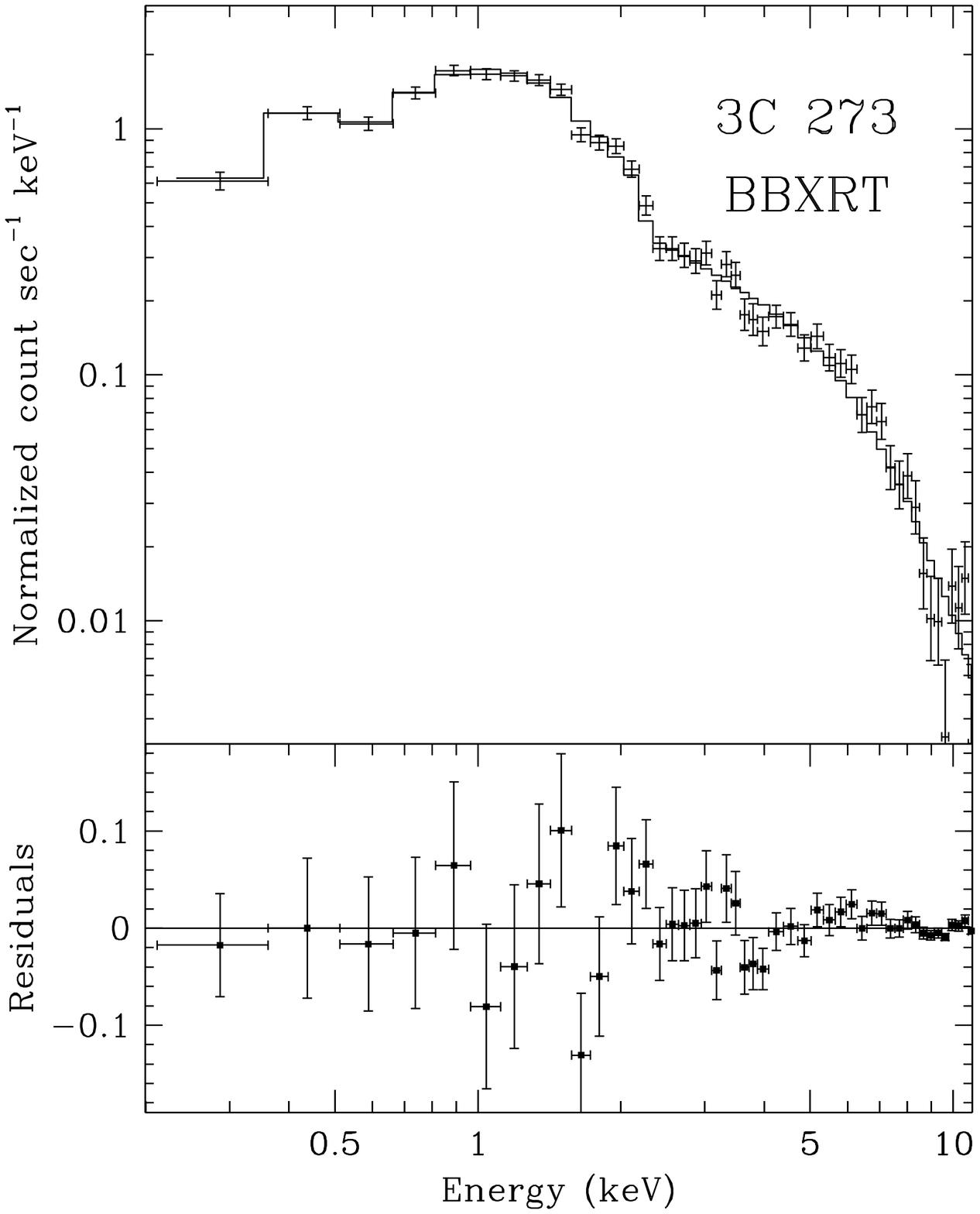}{4.15 in} {0}{50}{50}{-150}{-25}
\parbox{3.5in}{
\small\baselineskip 9pt
\footnotesize
\indent
{\sc Fig.}~5.---
{\it Upper panel:} the best-fit model to the Astro-1 BBXRT X-ray spectrum.
As described in \S4, the solid line is a dual power law model with Galactic
absorption and an empirical absorption edge at 0.5 keV folded through the
BBXRT response function. The error bars are 1$\sigma$.
{\it Lower panel:} residuals to the fit.
\label{fig5.ps}}
\vbox to 14pt{\vfill}
}

\vbox to 4.1in {
\begin{center}
\small
{\sc TABLE 4\\
Astro-1 Emission Lines\label{tbl-4}}
\vskip 2pt
\small
\begin{tabular}{lccc}
\hline
\hline
Feature & $\lambda_{obs}$ & Flux$\rm ^a$ & FWHM \\
                  & (\AA) &          	 & ($\rm km~s^{-1}$) \\
\hline
S~VI, $\lambda$ 933.38 & $1079.42 \pm 0.20$ & $\phantom{0}2.0 \pm 1.0$ & $\phantom{0}4905 \pm \phantom{0}721$ \\
S~VI, $\lambda$ 944.52 & $1092.32 \pm 0.20$ & $\phantom{0}1.0 \pm 0.5$ & $\phantom{0}4905 \pm \phantom{0}721$ \\
C~III, $\lambda$ 977.03 & $1129.95 \pm 0.20$ & $\phantom{0}3.8 \pm 1.2$ & $\phantom{0}4905 \pm \phantom{0}721$ \\
N~III, $\lambda$ 991.00 & $1146.12 \pm 0.20$ & $\phantom{0}2.9 \pm 1.1$ & $\phantom{0}4905 \pm \phantom{0}721$ \\
O~VI, $\lambda$1034.00 & $1191.46 \pm 0.20$ & $16.1 \pm 3.9$ & $\phantom{0}4225 \pm \phantom{0}776$ \\
O~VI, $\lambda$1034.00 & $1191.46 \pm 0.20$ & $13.7 \pm 5.5$ & $10554 \pm \phantom{0}723$ \\
O~VI, $\lambda$1034.00$\rm ^b$ & \nodata & $29.8 \pm 6.8$ & \nodata \\
No~ID & $1243.66 \pm 1.15$ & $\phantom{0}5.3 \pm 1.1$ & $\phantom{0}4905 \pm \phantom{0}721$ \\
C~III, $\lambda$1175.70 & $1360.28 \pm 0.20$ & $\phantom{0}6.6 \pm 1.8$ & $\phantom{0}4905 \pm \phantom{0}721$ \\
Ly$\alpha$, $\lambda$1215.67 & $1406.21 \pm 0.20$ & $56.3 \pm 3.8$ & $\phantom{0}2782 \pm \phantom{0}113$ \\
Ly$\alpha$, $\lambda$1215.67 & $1398.72 \pm 1.66$ & $49.8 \pm 5.1$ & $\phantom{0}8375 \pm \phantom{0}733$ \\
Ly$\alpha$, $\lambda$1215.67$\rm ^b$ & \nodata & $106.1 \pm 6.3$ & \nodata \\
N~V, $\lambda$1240.15 & $1431.80 \pm 0.52$ & $\phantom{0}6.1 \pm 1.8$ & $\phantom{0}2782 \pm \phantom{0}308$ \\
N~V, $\lambda$1240.15 & $1434.68 \pm 1.89$ & $41.0 \pm 3.8$ & $10554 \pm \phantom{0}723$ \\
N~V, $\lambda$1240.15$\rm ^b$ & \nodata & $47.1 \pm 4.2$ & \nodata \\
O~I, $\lambda$1304.35 & $1508.05 \pm 0.20$ & $\phantom{0}5.8 \pm 1.1$ & $\phantom{0}4905 \pm \phantom{0}721$ \\
C~II, $\lambda$1335.30 & $1544.34 \pm 0.20$ & $\phantom{0}2.1 \pm 1.2$ & $\phantom{0}4905 \pm \phantom{0}721$ \\
Si~IV, $\lambda$1402.77 & $1620.89 \pm 1.62$ & $15.1 \pm 2.4$ & $\phantom{0}6978 \pm 1325$ \\
C~IV, $\lambda$1549 & $1789.44 \pm 0.52$ & $17.8 \pm 2.5$ & $\phantom{0}2782 \pm \phantom{0}308$ \\
C~IV, $\lambda$1549 & $1793.03 \pm 1.89$ & $36.1 \pm 4.1$ & $10554 \pm \phantom{0}723$ \\
C~IV, $\lambda$1549$\rm ^b$ & \nodata & $53.9 \pm 4.8$ & \nodata \\
\hline
\end{tabular}
\vskip 2pt
\parbox{3.5in}{
\small\baselineskip 9pt
\footnotesize
\indent
$\rm ^a$$10^{-13}~\rm erg~cm^{-2}~s^{-1}$\\
$\rm ^b$Total flux for the narrow and broad components.
}
\end{center}
}

\vbox to 4.8in {
\begin{center}
\small
{\sc TABLE 6\\
Astro-1 Absorption Lines\label{tbl-6}}
\vskip 2pt
\small
\footnotesize
\begin{tabular}{lccc}
\hline
\hline
Feature & $\lambda_{obs}$ & $W_\lambda$ & FWHM$^{\rm a}$ \\
                  & (\AA) &  (\AA)      & ($\rm km~s^{-1}$) \\
\hline
O~I, $\lambda$ 948.69 & $\phantom{0}947.09 \pm 0.41$ & $\phantom{0}0.4 \pm 0.2$ & $\phantom{00}820 \pm \phantom{0}162$ \\
N~I~blend, $\lambda$ 953.00 & $\phantom{0}953.28 \pm 0.41$ & $\phantom{0}0.4 \pm 0.2$ & $\phantom{00}820 \pm \phantom{0}162$ \\
P~II,~N~I, $\lambda$ 964.00 & $\phantom{0}964.94 \pm 0.41$ & $\phantom{0}0.5 \pm 0.1$ & $\phantom{00}820 \pm \phantom{0}162$ \\
C~III, $\lambda$ 977.03 & $\phantom{0}975.68 \pm 0.30$ & $\phantom{0}0.7 \pm 0.1$ & $\phantom{00}820 \pm \phantom{0}162$ \\
S~III, $\lambda$1012.50 & \nodata & $<0.2$$\rm ^b$ & $\phantom{0}1147$ \\
Si~II, $\lambda$1020.70 & $1020.70 \pm 0.30$ & $\phantom{0}0.1 \pm 0.1$ & $\phantom{0}1162 \pm \phantom{0}128$ \\
O~VI, $\lambda$1031.93 & $1031.93 \pm 0.29$ & $\phantom{0}0.6 \pm 0.1$ & $\phantom{0}1162 \pm \phantom{0}128$ \\
O~VI,~C~II,~O~I, $\lambda$1038.00 & $1037.19 \pm 0.29$ & $\phantom{0}1.5 \pm 0.1$ & $\phantom{0}1344 \pm \phantom{0}148$ \\
Ar~I, $\lambda$1048.22 & $1049.04 \pm 0.39$ & $\phantom{0}0.5 \pm 0.1$ & $\phantom{0}1162 \pm \phantom{0}128$ \\
Fe~II, $\lambda$1063.18 & $1063.18 \pm 0.39$ & $\phantom{0}0.5 \pm 0.1$ & $\phantom{0}1162 \pm \phantom{0}128$ \\
Ar~I, $\lambda$1066.66 & $1067.51 \pm 0.39$ & $\phantom{0}0.5 \pm 0.1$ & $\phantom{0}1162 \pm \phantom{0}128$ \\
N~II, $\lambda$1084.19 & $1084.19 \pm 0.72$ & $\phantom{0}0.5 \pm 0.2$ & $\phantom{0}1162 \pm \phantom{0}128$ \\
Fe~II, $\lambda$1096.88 & \nodata & $<0.2$$\rm ^b$ & $779$ \\
Fe~II, $\lambda$1121.97 & $1121.97 \pm 0.72$ & $\phantom{0}0.3 \pm 0.1$ & $\phantom{00}779 \pm \phantom{0}127$ \\
N~I, $\lambda$1134.63 & $1134.96 \pm 0.72$ & $\phantom{0}0.4 \pm 0.2$ & $\phantom{00}779 \pm \phantom{0}127$ \\
Fe~II, $\lambda$1143.22 & $1142.81 \pm 0.72$ & $\phantom{0}0.4 \pm 0.1$ & $\phantom{00}779 \pm \phantom{0}127$ \\
Si~II~blend, $\lambda$1192.00 & $1191.56 \pm 0.72$ & $\phantom{0}1.1 \pm 0.3$ & $\phantom{0}1536 \pm \phantom{0}497$ \\
N~V, $\lambda$1238.82 & \nodata & $0.2$$\rm ^c$ & $536$ \\
N~V, $\lambda$1242.80 & \nodata & $0.1$$\rm ^c$ & $536$ \\
S~II, $\lambda$1250.50 & $1250.40 \pm 0.29$ & $\phantom{0}0.5 \pm 0.1$ & $\phantom{00}536 \pm \phantom{0}137$ \\
S~II,~Si~II, $\lambda$1260.00 & $1260.43 \pm 0.67$ & $\phantom{0}0.4 \pm 0.1$ & $\phantom{00}536 \pm \phantom{0}137$ \\
H~I, $\lambda$1275.19 & \nodata & $0.1$$\rm ^c$ & $536$ \\
C~II, $\lambda$1335.30 & $1335.19 \pm 0.34$ & $\phantom{0}1.0 \pm 0.2$ & $\phantom{00}964 \pm \phantom{0}176$ \\
H~I, $\lambda$1361.42 & \nodata & $0.1$$\rm ^c$ & $964$ \\
Ni~II, $\lambda$1370.09 & \nodata & $0.1$$\rm ^c$ & $589$ \\
Si~IV, $\lambda$1393.76 & $1393.35 \pm 0.51$ & $\phantom{0}0.2 \pm 0.1$ & $\phantom{00}589 \pm \phantom{0}100$ \\
Si~IV, $\lambda$1402.77 & $1402.31 \pm 0.51$ & $\phantom{0}0.1 \pm 0.1$ & $\phantom{00}589 \pm \phantom{0}100$ \\
Si~II, $\lambda$1527.17 & $1526.25 \pm 0.56$ & $\phantom{0}0.7 \pm 0.2$ & $\phantom{00}826 \pm \phantom{0}225$ \\
C~IV, $\lambda$1549.50 & $1548.73 \pm 0.26$ & $\phantom{0}1.3 \pm 0.2$ & $\phantom{00}715 \pm \phantom{0}137$ \\
Al~II, $\lambda$1670.79 & \nodata & $<0.4$$\rm ^b$ & $554$ \\
\hline
\end{tabular}
\vskip 2pt
\parbox{3.5in}{
\small\baselineskip 9pt
\footnotesize
\indent
$\rm ^a$All tabulated widths are consistent with the instrumental
resolution, and all features should be considered unresolved.\\
$\rm ^b$The upper limit for this line assumed a feature at the
nominal wavelength with a fixed FWHM as shown.\\
$\rm ^c$Our fits included a feature at the nominal wavelength with
the EW and FWHM fixed at the values shown.
}
\end{center}
}

\vbox to 4.1in {
\begin{center}
\small
{\sc TABLE 5\\
Astro-2 Emission Lines\label{tbl-5}}
\vskip 2pt
\small
\begin{tabular}{lccc}
\hline
\hline
Feature & $\lambda_{obs}$ & Flux$\rm ^a$ & FWHM \\
                  & (\AA) &          	 & ($\rm km~s^{-1}$) \\
\hline
S~VI, $\lambda$ 933.38 & $1079.59 \pm 0.15$ & $\phantom{0}4.2 \pm 0.2$ & $\phantom{0}5145 \pm \phantom{0}151$ \\
S~VI, $\lambda$ 944.52 & $1092.47 \pm 0.15$ & $\phantom{0}2.1 \pm 0.1$ & $\phantom{0}5145 \pm \phantom{0}151$ \\
C~III, $\lambda$ 977.03 & $1130.05 \pm 0.15$ & $\phantom{0}7.4 \pm 0.5$ & $\phantom{0}5145 \pm \phantom{0}151$ \\
N~III, $\lambda$ 991.00 & $1146.20 \pm 0.15$ & $\phantom{0}6.5 \pm 0.8$ & $\phantom{0}5145 \pm \phantom{0}151$ \\
O~VI, $\lambda$1034.00 & $1191.12 \pm 1.18$ & $13.6 \pm 9.2$ & $\phantom{0}2332 \pm \phantom{0}590$ \\
O~VI, $\lambda$1034.00 & $1191.12 \pm 1.18$ & $38.7 \pm 3.0$ & $10416 \pm \phantom{0}477$ \\
O~VI, $\lambda$1034.00$\rm ^b$ & \nodata & $52.3 \pm 9.6$ & \nodata \\
No~ID & $1243.66 \pm 1.15$ & $\phantom{0}5.6 \pm 0.8$ & $\phantom{0}5145 \pm \phantom{0}151$ \\
C~III, $\lambda$1175.70 & $1360.08 \pm 0.15$ & $10.0 \pm 1.0$ & $\phantom{0}5145 \pm \phantom{0}151$ \\
Ly$\alpha$, $\lambda$1215.67 & $1405.94 \pm 0.15$ & $70.7 \pm 3.3$ & $\phantom{0}2799 \pm \phantom{00}76$ \\
Ly$\alpha$, $\lambda$1215.67 & $1399.49 \pm 0.70$ & $93.5 \pm 4.4$ & $\phantom{0}8445 \pm \phantom{0}322$ \\
Ly$\alpha$, $\lambda$1215.67$\rm ^b$ & \nodata & $164.2 \pm 5.5$ & \nodata \\
N~V, $\lambda$1240.15 & $1433.61 \pm 0.40$ & $\phantom{0}6.6 \pm 1.6$ & $\phantom{0}3070 \pm \phantom{0}213$ \\
N~V, $\lambda$1240.15 & $1433.79 \pm 1.22$ & $52.8 \pm 3.5$ & $10416 \pm \phantom{0}477$ \\
N~V, $\lambda$1240.15$\rm ^b$ & \nodata & $59.4 \pm 3.8$ & \nodata \\
O~I, $\lambda$1304.35 & $1507.65 \pm 0.15$ & $\phantom{0}6.5 \pm 0.9$ & $\phantom{0}5145 \pm \phantom{0}151$ \\
C~II, $\lambda$1335.30 & $1543.90 \pm 0.15$ & $\phantom{0}5.3 \pm 1.3$ & $\phantom{0}5145 \pm \phantom{0}151$ \\
Si~IV, $\lambda$1402.77 & $1622.38 \pm 0.95$ & $21.5 \pm 2.0$ & $\phantom{0}6800 \pm \phantom{0}671$ \\
C~IV, $\lambda$1549 & $1791.30 \pm 0.40$ & $28.6 \pm 2.8$ & $\phantom{0}3070 \pm \phantom{0}213$ \\
C~IV, $\lambda$1549 & $1791.53 \pm 1.22$ & $57.8 \pm 3.2$ & $10416 \pm \phantom{0}477$ \\
C~IV, $\lambda$1549$\rm ^b$ & \nodata & $86.4 \pm 4.2$ & \nodata \\
\hline
\end{tabular}
\vskip 2pt
\parbox{3.5in}{
\small\baselineskip 9pt
\footnotesize
\indent
$\rm ^a$$10^{-13}~\rm erg~cm^{-2}~s^{-1}$\\
$\rm ^b$Total flux for the narrow and broad components.
}
\end{center}
}

\vbox to 5.4in {
\begin{center}
\small
{\sc TABLE 7\\
Astro-2 Absorption Lines\label{tbl-7}}
\vskip 2pt
\small
\footnotesize
\begin{tabular}{lccc}
\hline
\hline
Feature & $\lambda_{obs}$ & $W_\lambda$ & FWHM$\rm ^a$ \\
                  & (\AA) &    (\AA)    & ($\rm km~s^{-1}$) \\
\hline
O~I, $\lambda$ 948.69 & $\phantom{0}947.61 \pm 0.47$ & $\phantom{0}0.4 \pm 0.1$ & $\phantom{00}894 \pm \phantom{0}130$ \\
N~I~blend, $\lambda$ 953.00 & $\phantom{0}953.28 \pm 0.22$ & $\phantom{0}0.8 \pm 0.1$ & $\phantom{00}894 \pm \phantom{0}130$ \\
P~II,~N~I, $\lambda$ 964.00 & $\phantom{0}964.04 \pm 0.44$ & $\phantom{0}0.3 \pm 0.1$ & $\phantom{00}894 \pm \phantom{0}130$ \\
C~III, $\lambda$ 977.03 & $\phantom{0}976.48 \pm 0.15$ & $\phantom{0}1.0 \pm 0.1$ & $\phantom{00}894 \pm \phantom{0}130$ \\
Si~II, $\lambda$ 989.87 & $\phantom{0}989.31 \pm 0.15$ & $\phantom{0}0.5 \pm 0.1$ & $\phantom{00}894 \pm \phantom{0}130$ \\
N~III, $\lambda$ 991.00 & $\phantom{0}990.44 \pm 0.15$ & $\phantom{0}0.3 \pm 0.1$ & $\phantom{00}894 \pm \phantom{0}130$ \\
S~III, $\lambda$1012.50 & $1011.80 \pm 0.47$ & $\phantom{0}0.6 \pm 0.1$ & $\phantom{0}1202 \pm \phantom{0}129$ \\
Si~II, $\lambda$1020.70 & $1020.29 \pm 0.62$ & $\phantom{0}0.4 \pm 0.1$ & $\phantom{0}1202 \pm \phantom{0}129$ \\
O~VI, $\lambda$1031.93 & $1031.00 \pm 0.41$ & $\phantom{0}0.6 \pm 0.1$ & $\phantom{0}1202 \pm \phantom{0}129$ \\
O~VI,~C~II,~O~I, $\lambda$1038.00 & $1037.43 \pm 0.21$ & $\phantom{0}1.6 \pm 0.1$ & $\phantom{0}1386 \pm \phantom{0}149$ \\
Ar~I, $\lambda$1048.22 & $1051.70 \pm 0.69$ & $\phantom{0}0.5 \pm 0.1$ & $\phantom{0}1410 \pm \phantom{0}318$ \\
Fe~II, $\lambda$1063.18 & $1063.74 \pm 0.67$ & $\phantom{0}0.4 \pm 0.1$ & $\phantom{0}1410 \pm \phantom{0}318$ \\
Ar~I, $\lambda$1066.66 & \nodata & $<0.2$$\rm ^b$ & $1410$ \\
N~II, $\lambda$1084.19 & $1083.46 \pm 0.52$ & $\phantom{0}0.7 \pm 0.2$ & $\phantom{0}1410 \pm \phantom{0}318$ \\
Fe~II, $\lambda$1096.88 & \nodata & $<0.2$$\rm ^b$ & $767$ \\
Fe~II, $\lambda$1121.97 & $1121.89 \pm 0.46$ & $\phantom{0}0.2 \pm 0.1$ & $\phantom{00}767 \pm \phantom{0}126$ \\
N~I, $\lambda$1134.63 & $1134.83 \pm 0.18$ & $\phantom{0}0.7 \pm 0.1$ & $\phantom{00}767 \pm \phantom{0}126$ \\
Fe~II, $\lambda$1143.22 & \nodata & $<0.2$$\rm ^b$ & $767$ \\
Si~II~blend, $\lambda$1192.00 & $1191.33 \pm 0.76$ & $\phantom{0}1.9 \pm 1.8$ & $\phantom{0}1654 \pm \phantom{0}513$ \\
N~V, $\lambda$1238.82 & $1238.91 \pm 0.77$ & $\phantom{0}0.2 \pm 0.1$ & $\phantom{00}588 \pm \phantom{0}143$ \\
N~V, $\lambda$1242.80 & $1242.60 \pm 0.77$ & $\phantom{0}0.1 \pm 0.1$ & $\phantom{00}588 \pm \phantom{0}143$ \\
S~II, $\lambda$1250.50 & $1250.63 \pm 0.64$ & $\phantom{0}0.2 \pm 0.1$ & $\phantom{00}588 \pm \phantom{0}143$ \\
S~II,~Si~II, $\lambda$1260.00 & $1260.75 \pm 0.20$ & $\phantom{0}0.5 \pm 0.1$ & $\phantom{00}588 \pm \phantom{0}143$ \\
H~I, $\lambda$1275.19 & \nodata & $0.1$$\rm ^c$ & $588$ \\
C~II, $\lambda$1335.30 & $1334.34 \pm 0.21$ & $\phantom{0}0.7 \pm 0.1$ & $\phantom{00}651 \pm \phantom{0}103$ \\
H~I, $\lambda$1361.42 & \nodata & $0.1$$\rm ^c$ & $651$ \\
Ni~II, $\lambda$1370.09 & \nodata & $0.1$$\rm ^c$ & $599$ \\
Si~IV, $\lambda$1393.76 & $1393.68 \pm 0.17$ & $\phantom{0}0.5 \pm 0.1$ & $\phantom{00}599 \pm \phantom{0}100$ \\
Si~IV, $\lambda$1402.77 & $1402.63 \pm 0.17$ & $\phantom{0}0.2 \pm 0.1$ & $\phantom{00}599 \pm \phantom{0}100$ \\
No~ID, $\lambda$1492.00 & $1492.28 \pm 0.54$ & $\phantom{0}0.4 \pm 0.1$ & $\phantom{00}599 \pm \phantom{0}100$ \\
No~ID, $\lambda$1522.00 & $1522.85 \pm 0.25$ & $\phantom{0}0.2 \pm 0.1$ & $\phantom{00}246 \pm \phantom{00}92$ \\
Si~II, $\lambda$1527.17 & $1526.68 \pm 0.17$ & $\phantom{0}0.7 \pm 0.1$ & $\phantom{00}453 \pm \phantom{00}96$ \\
No~ID, $\lambda$1533.00 & $1531.85 \pm 0.25$ & $\phantom{0}0.5 \pm 0.1$ & $\phantom{00}453 \pm \phantom{00}96$ \\
C~IV, $\lambda$1549.50 & $1549.23 \pm 0.33$ & $\phantom{0}1.0 \pm 0.2$ & $\phantom{00}884 \pm \phantom{0}147$ \\
Al~II, $\lambda$1670.79 & $1670.24 \pm 0.42$ & $\phantom{0}0.6 \pm 0.2$ & $\phantom{00}539 \pm \phantom{0}170$ \\
\hline
\end{tabular}
\vskip 2pt
\parbox{3.5in}{
\small\baselineskip 9pt
\footnotesize
\indent
$\rm ^a$All tabulated widths are consistent with the instrumental
resolution, and all features should be considered unresolved.\\
$\rm ^b$The upper limit for this line assumed a feature at the
nominal wavelength with a fixed FWHM as shown.\\
$\rm ^c$Our fits included a feature at the nominal wavelength with
the EW and FWHM fixed at the values shown.
}
\end{center}
}

\vbox to 6.8in {
\vbox to 14pt{\vfill}
\plotfiddle{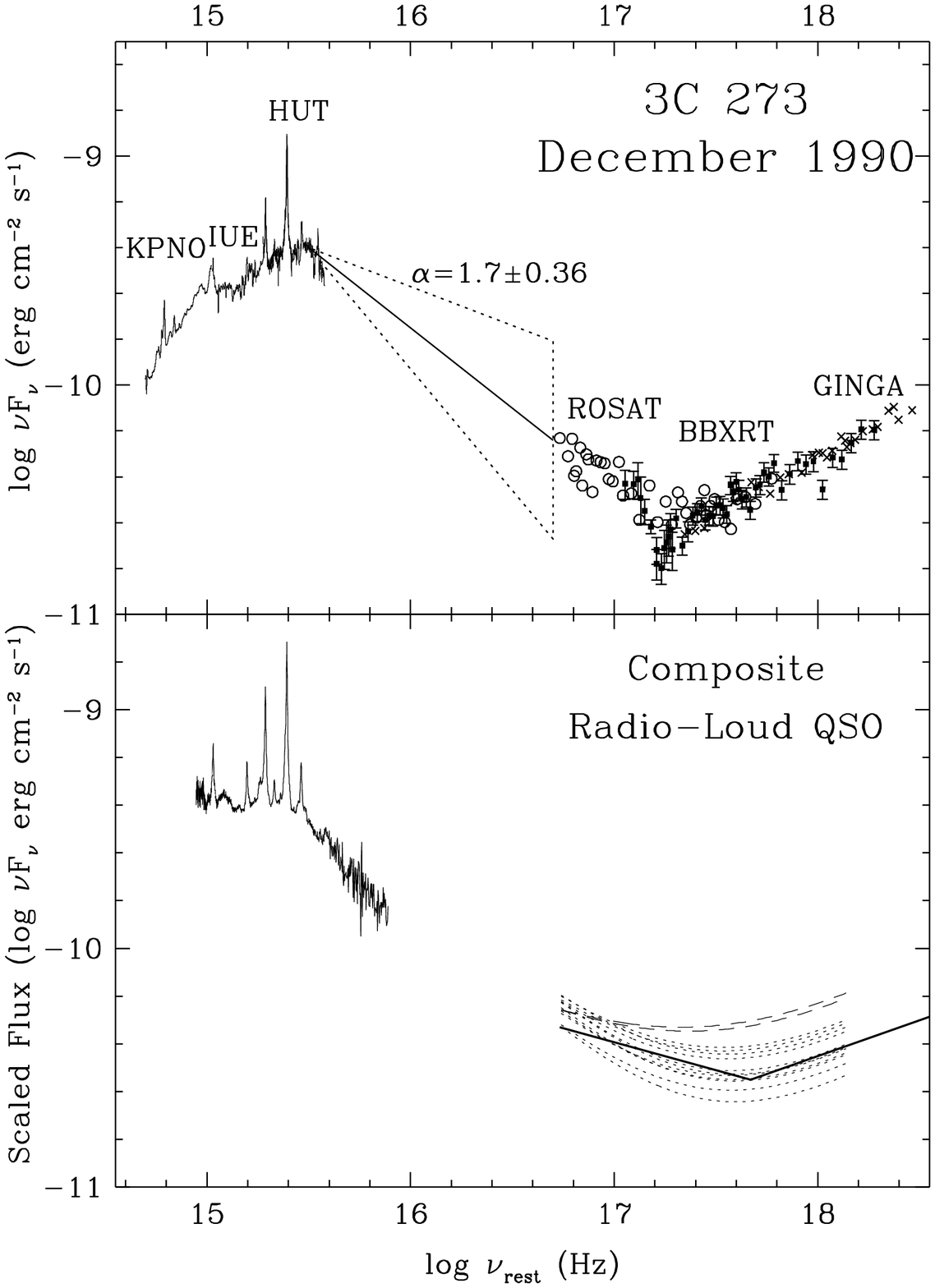}{4.45 in} {0}{50}{50}{-150}{-25}
\parbox{3.5in}{
\small\baselineskip 9pt
\footnotesize
\indent
{\sc Fig.}~6.---
{\it Upper panel:}
Quasi-simultaneous broad-band spectrum of 3C~273 from optical
to X-rays obtained in December 1990, corrected to the rest frame.
The ultraviolet and optical data have been corrected for reddening
with $E(B-V) = 0.032$, and the \ROSAT\ and BBXRT data have been corrected for
interstellar absorption with $N_H = 1.84 \times 10^{20}$ \cl.
To show the continuum shape more clearly, the UV and optical data have been
smoothed with a 9-pixel boxcar filter.
The best-fit power law for the short wavelength HUT data with $\alpha = 1.7$
is indicated, along with 1 $\sigma$ uncertainties for the slope.
{\it Lower panel:}
Composite radio-loud quasar spectrum in the UV-optical from
Zheng et al. (1997) and in the X-ray from Laor et al. (1997). The UV-optical
spectrum has been scaled to match the flux of 3C~273 at 1000 \AA, near the
break found in the composite spectrum (1050 \AA\ rest) and in the 3C~273
spectrum (919 \AA\ rest).
The X-ray spectrum has been scaled to the UV-optical composite using the mean
$\alpha_{ox}$ for the Laor et al. sample. The dotted and dashed curves
represent the spectral fits to 14 observations of 3C~273 over a 30-day period
with \ROSAT\ (LMP95). All but two of these observations (dashed) gave
$\alpha_{soft} = 1.64 - 1.78$.
\label{fig6.ps}}
\vbox to 14pt{\vfill}
}

Finally we have also plotted the results of an extensive set of observations of
3C~273 with \ROSAT\ in 1992 December to 1993 January by LMP95\markcite{LMP95}.
These authors obtained spectra at 2-day intervals over a period of a month.
They found that the only successful fits to their spectra were obtained
with a sum of two power laws, where they constrained the hard component to have
$\alpha_h = 0.5$ as determined by \EXOSAT\ and {\it Ginga} 2--10 keV
observations,
and also constrained the interstellar absorption at low energies to be
given by the Galactic value of $N_H = 1.84 \times 10^{20}$ \cl. Their soft
component was then found to have $\alpha_s = 1.7$ for all but two of their
observations, where they found $\alpha_s = 1.4$. We have plotted the fits
given in Table~6 of LMP95\markcite{LMP95} as dotted-line curves in
Fig.~6, with the two
discrepant observations as dashed-line curves. It is immediately apparent
that these observations of 3C~273 match quite well the composite radio-loud
X-ray spectrum of Laor et al. (1997)\markcite{Laor97} as we have scaled it to
match the UV-optical composite, which itself was scaled to match our
observation of 3C~273 at 1000 \AA.
They are also seen to extrapolate very well to match the
extreme ultraviolet part of our composite UV-optical spectrum (\cite{Zheng97}).

The upper panel of Fig.~6 thus demonstrates the ultraviolet peak in
the energy distribution of a single quasar, 3C~273, and a likely power-law
connection extending from the Lyman limit to the soft X-ray region, with
$\alpha_{EUV} = 1.7 \pm 0.36$,
though a substantial extreme ultraviolet gap necessarily remains unfilled.
The lower panel of Fig.~6, on the other hand, makes use
of composite quasar spectra constructed in the UV-optical and the X-ray
regions (suitably scaled) along with soft X-ray data for 3C~273
(LMP95\markcite{LMP95}), to
reveal a very similar spectrum, with a significantly smaller extreme
ultraviolet gap remaining unfilled.

The striking similarity between the 3C~273 spectrum and the spectral composite
provides genuine physical support for the reality of the composite
spectral shape derived by Zheng et al. (1997)\markcite{Zheng97}.
(We note that 3C~273 made no contribution to the shape of the composite
as only quasars with $z > 0.33$ were included in the sample used.)
One can argue that composite spectra in and of themselves are unphysical.
Zheng et al. (1997)\markcite{Zheng97} note that there is a wide dispersion in
spectral indices among individual quasars.
Koratkar \& Blaes (1999)\markcite{KB99} show several examples
of widely differing QSO spectral shapes.  So, there is no
guarantee that the composite assembled by Zheng et al. (1997)\markcite{Zheng97}
actually represents a real spectrum.  The fact that the spectrum of a single
object such as 3C~273 exhibits such remarkably similar characteristics
bolsters the case for the broad applicability of the composite spectrum.
Taken together, the composite spectrum and the 3C~273 spectrum
provide compelling evidence for a power-law spectrum from the Lyman limit to
the soft X-ray band in quasars, with a typical value of $\alpha = 1.7 - 2.2$.

\section{Accretion Disk Models}

To place our empirical results on the continuum spectral shape of 3C~273
in a physical context, we compare our spectra to accretion disk models.
Such models are well suited to producing spectra that peak in the far-UV,
and the spectral break near the Lyman limit in the quasar rest frame
suggests a mechanism related to the large change in opacity at the
intrinsic Lyman edge.
The presence of a break rather than an edge, however, plus the seeming
extrapolation of the high-frequency power law to soft X-ray energies are
suggestive of the appearance of an accretion disk spectrum with an
intrinsic Lyman edge feature that has been Comptonized by an external medium.
Comptonization (and relativistic effects in the disk) smear out any intrinsic
features in the disk spectrum as well as adding a power-law high-energy tail
(\cite{CZ91}).
As we will note below, however, the observed break is still sharper than
what can be accommodated by our simple, semi-empirical models.

To fit the broad-band spectral shape, we compute disk spectra in the
Schwarzschild metric that are a sum of blackbody spectra representing
rings in the disk running from an inner radius of 6 gravitational radii
($r_G =  G M_{BH} / c^2$) to 1000 $r_G$.
The blackbody spectra are modified by an empirical
Lyman-limit feature as described by Lee, Kriss, \& Davidsen
(1992)\markcite{LKD92} and Lee (1995)\markcite{Lee95}.
For the Comptonization we assume that a hot spherical medium with an
optical depth to Compton scattering of $\tau_e = 1$ surrounds the disk.
We use the formulation of Czerny \& Zbyszewska (1991)\markcite{CZ91} as
described by Lee, Kriss, \& Davidsen (1992)\markcite{LKD92} to calculate
the effects of the Comptonization.
The spectral index of the high energy tail in this calculation depends
almost entirely on the Compton $y$ parameter,
$y \propto \tau_e^2 T_c$, where $T_c$ is the temperature of the
Comptonizing medium (\cite{Lee95}), so the free parameters in our accretion disk
model are the mass accretion rate, $\dot{m}$, the mass of the central black
hole, $M_{BH}$, the optical depth at the Lyman edge, $\tau_{Ly}$,
the inclination of the disk, $i$, and $T_c$ (since we keep $\tau_e$ fixed at 1).

Our fits use the same wavelength intervals as the power-law and broken
power-law fits in \S3.2 and the same emission and absorption components.
All fits have the extinction fixed at $E(B-V)=0.032$.
Since there is little data in our HUT spectrum shortward of the spectral
break at $\sim$920 \AA\ in the rest frame, it is the slope and intensity
of the soft X-ray spectrum that largely determines $T_c$.
This has a best fit value of $T_c = 4.1 \times 10^8$ K in all our fits,
regardless of the other parameters.
The shape of the intrinsic disk spectrum is invariant for a fixed ratio
of $\dot{m} / M_{BH}^2$; this ratio is determined largely by the wavelength
of the peak in the spectral energy distribution. The normalization is
determined by the mass accretion rate $\dot{m}$. For conversion of flux to
luminosity, we assume a Hubble constant $H_0 = 75~\rm km~s^{-1}~Mpc^{-1}$
and a deceleration parameter $q_0 = 0$.
The strength of the spectral break in our model is related both to the peak in
the spectral energy distribution as well as
the optical depth at the Lyman edge, $\tau_{Ly}$.
The break wavelength and its sharpness constrain the best fit inclination $i$.
(Higher inclinations lead to
breaks at shorter wavelengths that are more smeared out.)

Our best fit to the Astro-1 spectrum,
for which simultaneous X-ray data and nearly simultaneous near-UV and
optical data are also available,
yields $\chi^2 = 1609$ for 1622 data points and 82 free parameters.
The best-fit parameters are summarized in the first column of Table~8.
(The emission and absorption line parameters have nearly identical values to
those presented from our previous empirical fits.)
The quoted errors are 90\% confidence for a single interesting parameter,
and they correspond to $\Delta \chi^2 = 4.6$.
The Comptonized accretion disk fits significantly better than a single power
law at the nominal value for Galactic extinction, but
worse than the best fitting broken power law.
The sharp break present in the spectrum is difficult to match with the smoother
shape of our simple accretion disk model, even with a Lyman-limit feature.

\begin{center}
\small
{\sc TABLE 8\\
Accretion Disk Models for 3C~273\label{tbl-8}}
\begin{tabular}{ l c c }
\hline
\hline
Parameter & Astro-1 & Astro-2 \\
\hline
$M_{BH}~(10^8~\Msun)$         & $7.1 \pm 0.3$       & $12 \pm 0.4$ \\
$\dot{m}~(\Msun~\rm yr^{-1})$ & $13 \pm 0.13$       & $12 \pm 0.12$ \\
$i~\rm (deg)$                 & $60^{+20}_{-10}$    & $60^{+25}_{-40}$ \\
$\tau_{Ly}$                   & $0.5 \pm 0.5$       & $0.0^{+0.2}_{-0.0}$ \\
$T_8~(10^8~\rm K)$            & $4.1 \pm 0.2$       & $4.1 \pm 0.2$       \\
$E(B-V)$                      & 0.032               & 0.032     \\
$\chi^2 / dof$                & 1609/1540           & 1752/1497 \\
\hline
\end{tabular}
\end{center}

The best fit to the Astro-2 spectrum is also summarized in Table~8.
The most significant difference between the Astro-1 and the Astro-2 fits
is that since the Astro-2 spectrum is redder overall, its fit favors
a higher mass black hole.
In all other respects, however, the Astro-1 and Astro-2 disk fits are similar.
Reflecting the nearly identical flux levels of the two observations, the mass
accretion rates are comparable.
Both fits favor an inclination $i = 60^\circ$, and the Lyman-limit
optical depth in each is low.
For Astro-1, $\tau_{Ly} = 0.5$, and $\tau_{Ly} = 0.0$ also gives an acceptable
fit; for Astro-2, $\tau_{Ly} = 0.0$.
Thus in the context of our simple accretion disk model a Lyman edge feature is
not formally required to achieve an acceptable fit to our data for either
Astro-1 or Astro-2. However, the fact that a broken power-law actually
provides a better fit, and that the break is very close to the Lyman
limit, hints that there probably is an effect associated with
an opacity change at the Lyman limit. We suggest that future more
sophisticated disk models should endeavor to match this feature.

Fig.~7 shows the best-fit accretion disk spectrum in comparison to
the HUT Astro-1 data and longer-wavelength points.
Here one can see that the rather smooth disk model does not fully
accommodate the sharp spectral break at $\sim$1000 \AA.
In the near-UV range the accretion disk continuum falls below the data
largely due to Balmer continuum and Fe~{\sc ii} line emission that is
not included in our model.
At longer wavelengths, the accretion disk falls below the data even more
as we have not included any emission components (either a power law
or dust emission) that are mainly responsible for the near-IR continuum.
Fig.~8 again shows the best fit over the whole frequency range from
the hard X-ray to 7000 \AA\ in the visible.
The Comptonized tail of the accretion disk spectrum provides an
excellent match to the soft X-ray excess present in the \ROSAT\
and the BBXRT data.

\vbox to 4.35in {
\vbox to 14pt{\vfill}
\plotfiddle{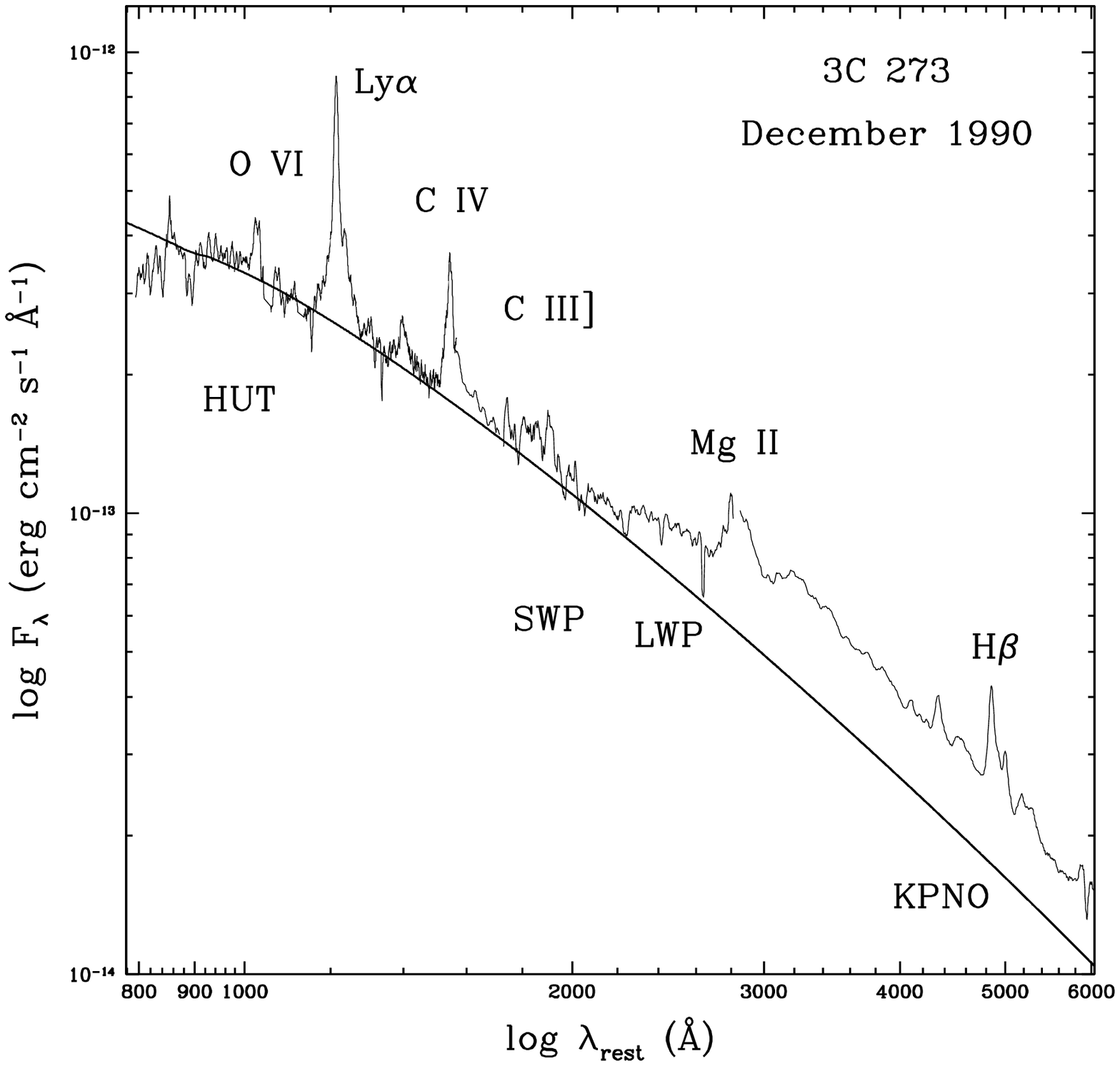}{3.21 in} {0}{47}{47}{-140}{-75}
\parbox{3.5in}{
\small\baselineskip 9pt
\footnotesize
\indent
{\sc Fig.}~7.---
Quasi-simultaneous UV-optical spectrum of 3C~273 from 1990
December, including data from HUT, IUE, and the KPNO 2.1 m telescope (kindly
provided by R. Green), corrected for reddening with $E(B-V) = 0.032$.
The data have been smoothed with a 9-pixel boxcar filter.
The smooth curve is the best-fitting accretion disk model as described in
Section 5, with parameters given in Table~8.
\label{fig7.ps}}
\vbox to 14pt{\vfill}
}

Several aspects of our accretion disk fits illustrate the shortcomings of
our simple model.
Our best-fit inclinations of $i = 60^{\circ}$
are a bit high given the superluminal jet in 3C~273 that suggests the
disk normal should be directed close to our line of sight.
However, inclinations in superluminal sources can be as high as $30-45^\circ$,
and this provides an acceptable fit for our Astro-2 observations.
An inclination of $0^\circ$ is excluded at high confidence in both data sets
($\Delta \chi^2 = 27$ for the Astro-1 data).
Also, as mentioned earlier, the inclination is determined mostly by the
wavelength of the spectral break.
A Kerr disk would produce bluer breaks that are more
smeared out at much lower inclination than in a Schwarzschild disk
(e.g., see \cite{LN89}).

\setcounter{figure}{7}
\begin{figure*}[t]
\plotfiddle{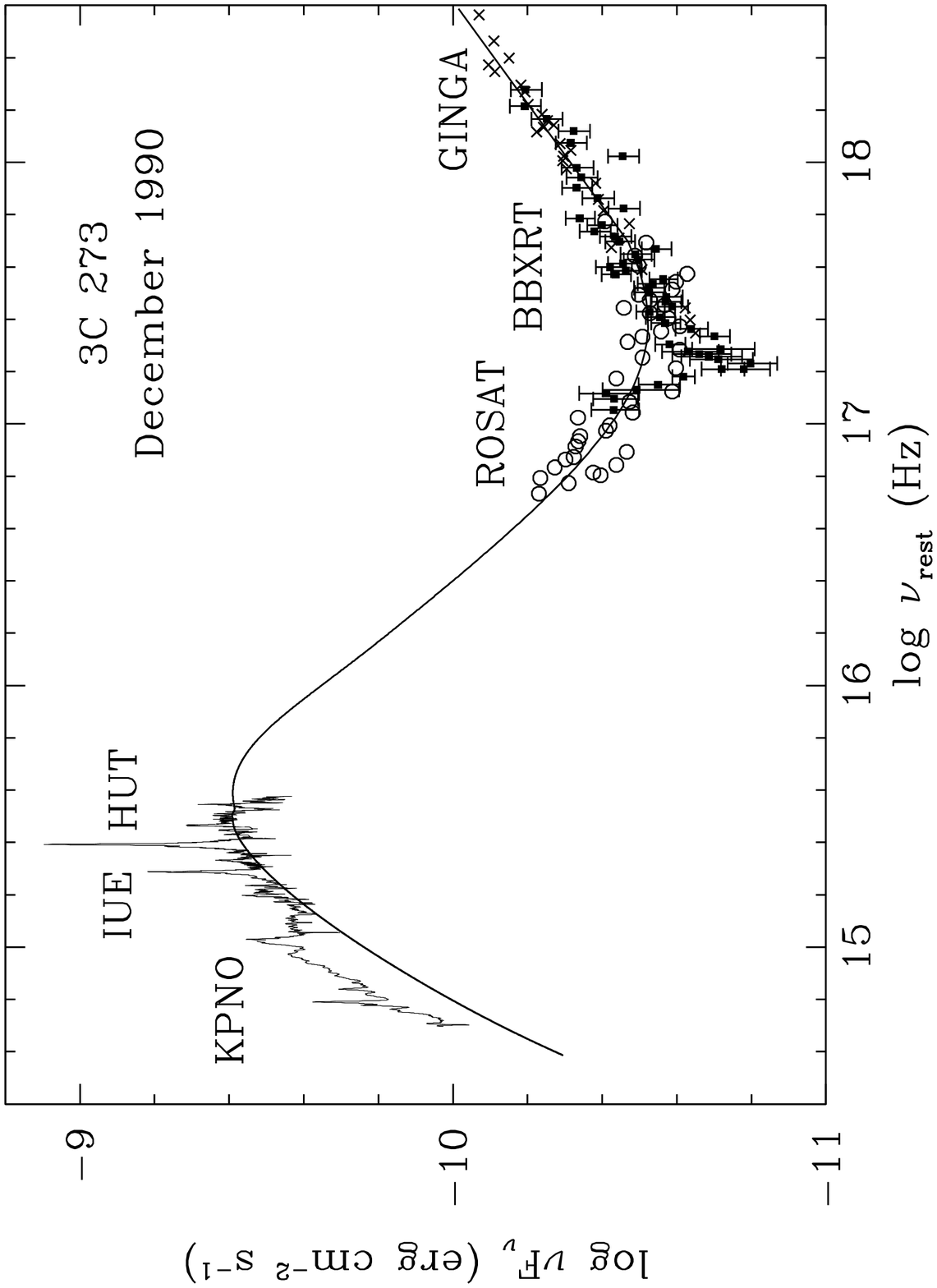}{5.0 in}{-90}{70}{70}{-275}{405}
\caption{
Quasi-simultaneous broad-band spectrum of 3C~273 as described
in Fig.~5, with the best-fitting Comptonized accretion disk model
added to a hard X-ray power-law with $\alpha = 0.5$.
The sharp turn downward of the observed data below the model at the shortest
ultraviolet wavelengths is simply absorption by the converging hydrogen Lyman
series, and it is fully accounted for in our models (see Fig.~3).
\label{fig8.ps}}
\end{figure*}

Both the Astro-1 and Astro-2 fits have accretion rates that exceed the
thin-disk limit of $L / L_{Edd} \sim 0.3$ for a Schwarzschild metric.
For the Astro-1 fit, $L / L_{Edd} = 0.46$; for Astro-2, $L / L_{Edd} = 0.25$.
A Kerr metric would also alleviate this problem: the limit for a thin disk in
this metric is higher, and, at the lower inclinations it would require, the
accretion rate would also be lower.

Finally, while the Astro-1 and Astro-2 fits have similar mass accretion rates,
their best-fit black hole masses differ by nearly a factor of two.
This illustrates a fundamental difficulty of simple steady-state
accretion disk models in dealing with variability:
it occurs on time scales more rapid than the viscous time
scale that governs the applicability of steady-state models.
In a steady-state model for a given object, the
only variable that can vary freely is the mass accretion rate.
Therefore, one should observe correlated changes in flux and color, with
brighter states exhibiting bluer disk spectra.
However, the Astro-1 and Astro-2 data are nearly identical in flux,
but they have significantly different colors.
Without a good model for accommodating variability in disk models,
one can see that factors of several uncertainty in the actual physical
parameters are easily present in our results.
It is reassuring to note, however, that our values for the black hole
mass based on accretion disk models ($7.1 - 12 \times 10^8~\Msun$) bracket the
independent value of $7.4 \times 10^8~\Msun$ obtained by
Laor (1998)\markcite{Laor98} using reverberation-mapping data.

We conclude that our accretion disk models of the Astro-1 and Astro-2 spectra
provide an overall qualitative physical characterization of the 3C~273 spectrum.
While the models have quantitative shortcomings, they provide an empirical
guide for potentially more sophisticated models.
The most important aspects of our simple description in terms of
blackbody emission and Comptonization are that it
(1) accounts for the peak of the ultraviolet spectrum, (2) produces a
spectral break in the far-UV, and (3) has a hard power-law tail that extends
to the soft-X-ray band and matches the observed soft-X-ray spectrum.

\section{Summary}

We have presented absolutely-calibrated ultraviolet
spectrophotometry over the 900--1800 \AA\ range for the quasar 3C~273, obtained
with the Hopkins Ultraviolet Telescope on the Astro-1 mission in December
1990 and on the Astro-2 mission in March 1995. In both observations the
continuum displays a change of slope near the Lyman limit in the
quasar rest frame. At longer UV wavelengths the continuum is
well-represented by a power-law 
of $\alpha_1 = 0.5$.
Shortward of the Lyman limit, however, the continuum slope has $\alpha_2 =
1.7 \pm 0.36$, where the uncertainty includes our uncertainty about $E(B-V)$
at the level of $\pm 0.01$.

The energy distribution per logarithmic frequency interval $\nu f_\nu$
therefore has a peak close to the quasar Lyman limit. The short wavelength
extreme UV power-law extrapolates very well to match the soft x-ray
spectrum of 3C~273 obtained nearly simultaneously in the case of Astro-1
with BBXRT and \ROSAT. The soft x-ray data themselves give
$\alpha_s=1.7(\pm 0.1)$ (LMP95\markcite{LMP95}), so the combined UV and X-ray
data suggest $\alpha_{UV-X} = 1.7$. While some models for the photoionizing
extreme ultraviolet radiation from quasars have a peak --- the so-called
``big blue bump'' --- at a wavelength much shorter than the Lyman limit at
912 \AA\ with a peak flux much higher than what is actually observed at both
longer and shorter wavelengths (e.g., \cite{MF87}; \cite{Bechtold87}; 
\cite{Gondhalekar92}), 
we find that in 3C~273 the peak of the big blue bump occurs very close to the
Lyman limit.
The photoionizing flux from 3C~273 is given
by $f_\nu = f_{LL} (\nu/\nu_{LL})^{-1.7}$ extending from the Lyman limit to
1 keV, although there is still a gap in the data covering one decade of
frequency in the extreme ultraviolet.

Analysis of composite quasar spectra constructed in the ultraviolet
(\cite{Zheng97}) and the X-ray (\cite{Laor97}) bands leads to the same result.
The similarity between the spectrum of a single object such as 3C~273 and
these composites lends physical credence to their applicability.
It therefore appears to be true of quasars in general that the peak
of the continuum energy distribution occurs near the Lyman limit, and that
the ionizing continuum is well-represented  by a power-law of energy index
$\alpha_{UV-X} = 1.7 - 2.2$ extending to about 1 keV, where a separate hard
X-ray component begins to dominate for higher energies. The ``extreme
ultraviolet gap'' in the composite quasar spectrum is now only half a decade
in frequency, however. This can be reduced somewhat further by extending
the work of Zheng et al. (1997)\markcite{Zheng97} to include more observations
of high redshift quasars in the far ultraviolet band.

The general shape of the optical, ultraviolet, and soft x-ray
spectrum of 3C~273 is consistent with that of an optically thick accretion
disk around a massive black hole, which is itself surrounded by a hot
medium that modifies the emergent spectrum through the inverse Compton
effect, as suggested by Czerny \& Zbyszewska (1991)\markcite{CZ91}, Lee et al.
(1992)\markcite{LKD92} and Lee (1995)\markcite{Lee95}.
In our model the spectral break is due to the thermal peak
of the accretion disk spectral energy distribution with a minor contribution
from Lyman edge absorption.
Both features are broadened and blurred by Comptonization in a surrounding
hot medium, perhaps a corona or wind emanating from the disk.
Comptonization of the thermal photons from the disk also produces the
power-law tail extending to the soft X-ray region (e.g., \cite{CE87};
\cite{MM90}; \cite{RFM92}).
Our simple model is unable to fully account for the sharpness of the spectral
break, however, and we suggest that a more sophisticated treatment of the
Lyman-edge region may eventually provide a better match to these data.

Assuming a Schwarzschild black hole with an inclination of 60 degrees,
the UV spectrum is fit with a black hole mass $M_{bh} = 7 \times 10^8~\Msun$
with an accretion rate $\dot{M} = 13~\Msun/yr$.
Superposed on the disk spectrum is an empirically determined Lyman edge
with $\tau_{Ly} = 0.5$. 
The hot corona or wind is required to have a Compton parameter $y \approx 1$,
which is obtained, for example, with $\tau_{es} = 1$
and $T_e = 4 \times 10^8$ K.
A temperature of this order appears plausible for a corona where Compton
cooling balances Compton heating by the hard x-ray flux in 3C~273. These
results are quite similar to those we found previously by fitting similar
models to a composite quasar spectrum (\cite{Zheng97}). There we found,
assuming a typical inclination of 30 degrees, a black hole mass
$M_{bh} = 1.4 \times 10^9~\Msun$ and
an accretion rate $\dot{M} = 2.8~\Msun/yr$.
While the
best-fit parameters would change somewhat in the more realistic case of a
rotating (Kerr) black hole (which we have not calculated), we believe these
results provide strong evidence in favor of the hypothesis that quasars are
powered by accretion onto massive black holes.

\acknowledgments

We thank R. Green for providing the optical data, J. Kruk and C. Bowers
for help with 
the Astro-1 data, and K. Weaver for assistance with the X-ray data reduction.
The Hopkins Ultraviolet Telescope project has been supported by NASA contract
NAS--5-27000 to the Johns Hopkins University.

\end{document}